\documentclass[a4paper,11pt]{article}
\pdfoutput=1
\usepackage{jheppub}
\usepackage[T1]{fontenc}
\usepackage{amsfonts,amssymb,epsfig,verbatim,mathbbol,amstext,amsmath,psfrag}
\usepackage{graphicx}
\usepackage{t1enc}
\usepackage{subfigure}
\usepackage{dsfont}
\usepackage{slashed,MnSymbol}

\usepackage{colordvi}
\usepackage{xcolor}
\usepackage{color}

\newcommand{\ev}[1]{\left\langle #1 \right\rangle}
\newcommand{\cref}[1]{(\ref{#1})}

\newcommand{\be}{\begin{equation}}
\newcommand{\ee}{\end{equation}}

\newcommand{\Z}{\mathcal{Z}}

\newcommand{\mui}{\mu_I}

\title{Equation of state and speed of sound of isospin-asymmetric QCD on the lattice}

\author[a]{B.~B.~Brandt}
\author[b]{F.~Cuteri}
\author[a]{G.~Endr\H{o}di}

\affiliation[a]{Fakult\"at f\"ur Physik, Universit\"at Bielefeld,
D-33615 Bielefeld, Germany}
\affiliation[b]{Institut f\"ur Theoretische Physik, Goethe Universit\"at Frankfurt, D-60438 Frankfurt am Main, Germany}

\emailAdd{brandt@physik.uni-bielefeld.de}
\emailAdd{cuteri@itp.uni-frankfurt.de}
\emailAdd{endrodi@physik.uni-bielefeld.de}

\abstract{
We determine the QCD equation of state at nonzero temperature in the presence of an isospin asymmetry between the light quark chemical potentials on the lattice. Our simulations employ $N_f=2+1$ flavors of dynamical staggered quarks at physical masses, using three different lattice spacings.
The main results,
obtained at the individual lattice spacings,
are based on a two-dimensional spline interpolation of the isospin density,
from which all relevant quantities can be obtained analytically.
In particular, we present results for the pressure,
the interaction measure, the energy and entropy densities, as well as the speed of
sound.
Remarkably, the latter is found to exceed its ideal gas limit deep in the pion condensed phase,
the first account of the violation of this limit in first principles QCD.
Finally, we also compute the phase diagram in the temperature
-- isospin density plane for the first time.
Even though the results are not continuum extrapolated and thus not final,
the data for all observables will be useful for the benchmarking of effective theories
and low-energy models of QCD and are provided in ancillary files for simple reuse.
}

\begin{document} 
\maketitle
\flushbottom

\section{Introduction}

The theory of the strong interactions is Quantum Chromodynamics (QCD), 
featuring confinement of quarks and gluons at low 
energies, as well as asymptotic freedom at high scales. 
Albeit radically different in their properties, these two phases of strongly interacting matter are connected by a smooth crossover 
transition at zero net quark density according to lattice QCD simulations~\cite{Aoki:2006we,Bhattacharya:2014ara}. 
How the dominant degrees of freedom transform 
from composite objects (hadrons) to colored quarks and gluons through this transition is described by the equation of state (EoS) of the system.
In particular, the EoS gives a complete description of equilibrium QCD in terms of 
a relationship between thermodynamic observables including the pressure, the energy
density or the entropy density. The phenomenological relevance of the EoS is manifold and ranges from heavy-ion physics to astrophysics and cosmology. The above observables control the evolution of the quark-gluon plasma in hydrodynamic models of heavy-ion collisions~\cite{Teaney:2001av,
Kolb:2003dz}, the expansion of the early universe via the Friedmann equations~\cite{Boyanovsky:2006bf} and also the mass and radius of stable neutron stars through the Tolman-Oppenheimer-Volkoff equation~\cite{Lattimer:2000nx}. In the latter context, a particularly relevant feature of the EoS is the speed of sound $c_s$ of QCD matter and the related polytropic index $\gamma$, which 
might serve as a proxy to distinguish stars with and without deconfined quark matter cores~\cite{Tews:2018kmu,Annala:2019puf}.

The above physical systems contain QCD matter in very different 
environments and therefore require the knowledge of the EoS as 
a function of different control parameters. These parameters include 
the temperature $T$, the chemical potentials $\mu$ conjugate to the conserved charges, as well as further variables like external electromagnetic fields.
The relevant chemical potentials are charge $\mu_Q$, baryon $\mu_B$ and strangeness $\mu_S$
chemical potentials. 
While the baryon chemical potential is in most cases assumed to carry the dominant effect,
in some cases $\mu_Q$ can play the major role.
This occurs for example for an early Universe featuring large lepton flavour asymmetries~\cite{Oldengott:2017tzj,Wygas:2018otj,Middeldorf-Wygas:2020glx}. Here the isentropic cosmological expansion leads to substantial charge chemical potentials, triggering the onset of pion condensation and producing characteristic signals for primordial black holes and gravitational wave spectra~\cite{Vovchenko:2020crk}.
Significant negative charge chemical potentials also arise for QCD 
matter in neutron stars due to an excess of down quarks over up quarks.

\begin{figure}[t]
 \centering
\vspace*{-2mm}
\includegraphics[width=.48\textwidth]{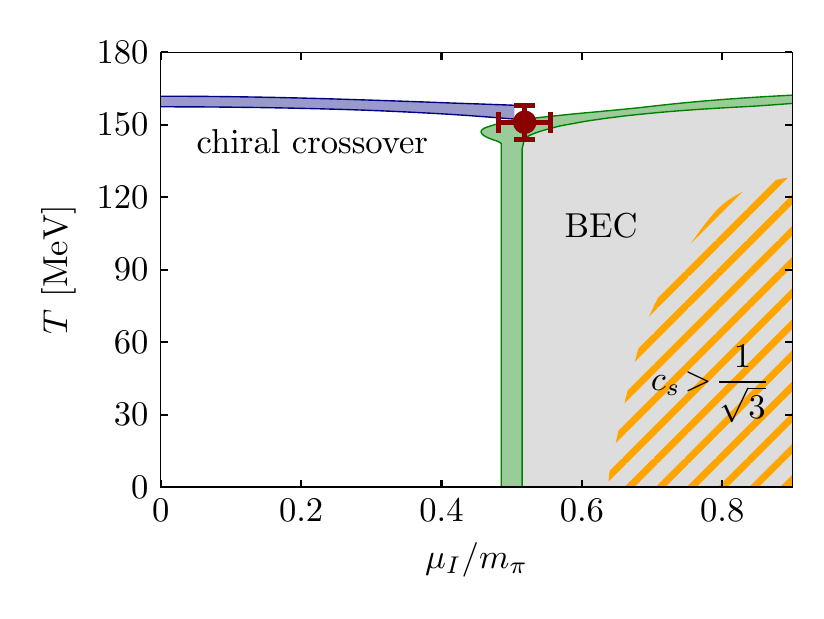}
 \caption{\label{fig:phd}
 Phase diagram of isospin asymmetric QCD determined in Ref.~\cite{Brandt:2017oyy}.
 The green band is the phase boundary to the phase with Bose-Einstein condensation (BEC),
 the blue band is the prolongation of the chiral crossover in the $T$-$\mu_I$ plane
 and the red data point marks the pseudo-triple point, the meeting point of the
 crossover and the BEC phase. In yellow we also show a sketch of the region,
 where the speed of sound exceeds its conformal limit.}
\end{figure}

The charge chemical potential can be rewritten in terms of a nonzero isospin chemical potential $\mu_I$. While lattice simulations with a generic combination of quark chemical potentials suffer from the infamous complex action (or sign) problem, QCD at pure isospin chemical potential, i.e., at vanishing other chemical potential components, has a real action and is amenable to direct Monte-Carlo simulations~\cite{Son:2000xc}. 
At low temperature, this setting exhibits a second-order phase transition to a phase with a Bose-Einstein condensate (BEC) of charged pions according to chiral perturbation theory~\cite{Son:2000xc}. First lattice simulations with higher-than-physical quark masses have qualitatively confirmed this expectation~\cite{Kogut:2002tm,Kogut:2002zg,Kogut:2004zg,Endrodi:2014lja} and also gave important insight to the structure of the phase diagram in the $T$-$\mu_I$ plane~\cite{deForcrand:2007uz,Cea:2012ev} as well as pion dynamics at low $T$~\cite{Detmold:2012wc}. 
In Refs.~\cite{Brandt:2017oyy,Brandt:2018omg} we carried out a systematic 
investigation of this system with physical quark masses and determined the phase diagram in the continuum limit, revealing an interesting interplay of chiral symmetry breaking, deconfinement and Bose-Einstein condensation. The continuum phase diagram for the parameter space relevant for this
study is shown in Fig.~\ref{fig:phd}.

In this paper we determine the EoS throughout the phase diagram for 
a broad range of temperatures and isospin chemical potentials. Generalizing our approach at (approximately) zero temperature~\cite{Brandt:2018bwq}, we construct the pressure, the energy and entropy densities, the interaction measure and the speed of sound from the isospin density as primary observable.
At low temperatures and high $\mu_I$, we find that the speed of sound increases above its conformal limit $1/\sqrt{3}$ (we use natural units, with the speed of light
set to unity).
This is the first evidence
for the explicit violation in first principles QCD of this
general bound expected from holography~\cite{Cherman:2009tw} and, together with the polytropic
index, which we compute as well,
might provide relevant information for the modeling of the EoS based 
on neutron star radii and masses~\cite{Tews:2018kmu,Annala:2019puf}.
We also included a sketch of the region where $c_s$ exceeds this conformal bound
in Fig.~\ref{fig:phd}.
Besides the EoS, we also use our results to draw the QCD phase diagram in the temperature -- isospin density plane. This result, together 
with the complete tabulated EoS as shown in the plots, is available in the ancillary files
submitted to the arXiv along with the preprint of this paper. To facilitate the use of the EoS in phenomenological models, we provide an accompanying data publication~\cite{datapub}, including the physical observables and the uncertainties for all temperatures and chemical potentials where results are available. The results can be used as benchmarks for low-energy models and effective theories of QCD as well as for comparing to functional approaches.
First accounts of our findings have been given in Refs.~\cite{Vovchenko:2020crk,Brandt:2017zck,Brandt:2018wkp,Brandt:2021yhc}.

\section{Determination of the EoS}
\label{sec:comp}

\subsection{Simulation setup and main observables}
\label{sec:simobs}

In our simulations we use
$N_f=2+1$ flavors of rooted staggered quarks with two steps of stout smearing
at physical quark masses and the
tree-level Symanzik improved gluon action. The line of constant physics for the bare quark masses is taken from
Ref.~\cite{Borsanyi:2010cj}. 
For the approach to the continuum limit
we use lattices with temporal extents $N_t=8,\,10$ and $12$ and
aspect ratios of $N_s/N_t\approx 3$ (these ensembles have already been used for the phase diagram~\cite{Brandt:2017oyy}), together with a number of additional ensembles at $T=0$.
More details concerning the run parameters are collected in appendix~\ref{app:sim-detail}.
The isospin chemical potential $\mu_I$ enters the light quark Dirac operator in an exponential
form and is normalized such that pion condensation sets in at zero temperature at $\mu_I=m_\pi/2$.
As in
our previous studies~\cite{Brandt:2017oyy,Brandt:2018omg,Brandt:2018bwq,Vovchenko:2020crk} the simulations are performed including a pionic source parameter $\lambda$,
in the light quark mass matrix (see also
Refs.~\cite{Kogut:2002tm,Kogut:2002zg,Endrodi:2014lja}), 
which serves as an infrared regulator and triggers pion condensation 
in a finite volume.
Physical results are obtained by means of an extrapolation $\lambda\to0$, which 
is facilitated by improving the observables and reweighting the configurations.
For details on this improvement, see Refs.~\cite{Brandt:2017oyy,Brandt:2018omg,Brandt:2018bwq}. 
For computing uncertainties we use the bootstrap procedure with 1000 samples.

As we will see below, the main observable is the isospin
density,
\be
\label{eq:nI-def}
n_I = \frac{T}{V}\frac{\partial \log\Z}{\partial \mu_I}\,,
\ee
from which the full $\mu_I$-dependence of the EoS can
be extracted. The strategy for the EoS computation will be
outlined below. $n_I$ can be computed directly from the
simulations as described in Ref.~\cite{Brandt:2018bwq}. To perform
the $\lambda$-extrapolations we use the improvement program
introduced in Ref.~\cite{Brandt:2017oyy} with the application
to $n_I$ as explained in Ref.~\cite{Brandt:2018bwq,Brandt:2018omg}.
This improvement program results in fully controlled
extrapolations and from now on we only discuss results which
have already been extrapolated to $\lambda=0$. We note
that a well controlled
$\lambda$-extrapolation is of particular importance to facilitate the following spline interpolations of $n_I$ in $T$ and in $\mu_I$.

\subsection{The EoS from an interpolation of the isospin density}
\label{sec:eos-int}

Apart from the isospin density $n_I$, the main task for the
determination of the EoS is the computation of the pressure $p$ and
the interaction measure $I$. All other relevant quantities, apart from the
speed of sound (to be discussed in Sec.~\ref{sec:cs-comp}) follow from these three
quantities. In particular, the energy and entropy densities are
given by
\be
\label{eq:ene-ent}
 \epsilon = I+3p, \quad
 \quad s=\frac{\epsilon + p - \mu_I n_I}{T} \,.
\ee

The EoS at vanishing chemical potential has been computed in the
continuum limit in various setups and by different
collaborations, see e.g.\ Refs.~\cite{Borsanyi:2013bia,HotQCD:2014kol}.
It is thus convenient to separate the effects due to nonzero temperature from
the modifications due to the presence of a nonzero isospin chemical potential.
This is possible for the quantities of this section, for which the two types of
contributions are added, but not for the speed of sound, defined via directional
derivatives in the $T$-$\mu_I$ plane. The pressure and the interaction measure can
be written as
\be
\label{eq:eos-decomp}
\begin{array}{rcl} \displaystyle p(T,\mu_I) & = & \displaystyle p(T,0) + \Delta p(T,\mu_I)\,,
\vspace*{2mm} \\
\displaystyle I(T,\mu_I) & = & \displaystyle I(T,0) + \Delta I(T,\mu_I) \,,
\end{array}
\ee
where the modifications of the EoS due to the isospin chemical potential,
$\Delta p$ and $\Delta I$, are the objects of interest in our study. Whenever
we need to use results for $p(T,0)$ and $I(T,0)$, we use the results obtained
from a reanalysis of the data of Ref.~\cite{Borsanyi:2013bia} with the
parameterisation of $I$ discussed in section 3 of Ref.~\cite{Borsanyi:2010cj}.
The correct inclusion of the correlations of the associated parameters are of
particular relevance for the computation of the speed of sound. The details
of the $\mu_I=0$ data are discussed in appendix~\ref{app:mui0-eos}.

A possible starting point for the computation of the EoS is the relation
\be
\label{eq:intmeas}
 \frac{I(T,\mu)}{T^4} = T \frac{\partial}{\partial T} \left(\frac{p(T,\mu_I)}{T^4}\right)
+ \frac{\mu_I n_I(T,\mu_I)}{T^4} \,.
\ee
At vanishing chemical potential this equation is used to rewrite $I$
as a derivative of the partition function with
respect to the lattice scale~\cite{Engels:1990vr,Blum:1994zf,Engels:1996ag,Aoki:2005vt}.
At nonzero chemical potential, one can follow a similar strategy to calculate the
modifications of pressure and interaction measure using $\mu=0$ subtraction (see, e.g.,
Ref.~\cite{Allton:2003vx} and Refs.~\cite{Iida:2022hyy,Itou:2022ebw}).
The direct application of this method at nonzero $\mu_I$
is discussed in appendix~\ref{app:eos-icomp}, where we will see that it leads
to large uncertainties for the interaction measure, due to the
$\mu_I=0$ subtractions. An alternative, which leads to more accurate results that we will
present in the following, is to use a two-dimensional smooth
interpolation of the results for $n_I$ to obtain the function $n_I(T,\mu_I)$.\footnote{We note that in the infinite volume, $n_I(T,\mu_I)$ has a kink at the second order phase
transition to the BEC phase, which cannot be described by a smooth function.
Here we are working in a finite volume, where this kink is absent; it only reappears
in the thermodynamic limit. Therefore a description of this
region using smooth splines is possible.} Since the
isospin density is the derivative of the pressure with respect to $\mu_I$,
see Eq.~\eqref{eq:nI-def}, the modification of the pressure can be computed from such an
interpolation as~\cite{Vovchenko:2020crk,Brandt:2018wkp,Brandt:2017zck}
\be
\label{eq:plat}
 \Delta p(T,\mu_I) = \int_0^{\mu_I} d\mu'_I \, n_I(T,\mu'_I) \,.
\ee
Inserting this into Eq.~\eqref{eq:intmeas} gives for the interaction measure (see also Ref.~\cite{Vovchenko:2020crk})
\begin{equation}
\label{eq:Ilat}
 \Delta I(T,\mui) = \mui n_I(T,\mui) + \int_0^{\mui} d\mui'\, \Big[ T \frac{\partial}{\partial T} - 4 \Big] n_I(T,\mui') \,.
\end{equation}

The remaining task is to perform the interpolation and to obtain $n_I(T,\mu_I)$ as a
two-dimensional function.
Since the interpolation of an unknown function based on a discrete set of points subject
to statistical uncertainties is an ill-posed inverse problem, the final 
interpolation will not be unique. This is already true for the interpolations used for
the computation of the EoS at $\mu=0$ and will extend to the computation of the modifications of the
EoS due to nonzero $\mu_I$ discussed in the next section. The task at hand is to obtain
an interpolation which is close to the actual physical solution while remaining as
model-independent as possible. For the purpose of model-independence we average over
all possible two-dimensional cubic spline interpolations with variable spline nodepoints
(spline fits), weighted with the goodness of the description of the data. Here the
goodness of the description is determined via the Akaike information
criterion~\cite{Akaike1973InformationTA} and we have included a term to suppress unwanted
(and unphysical) oscillatory solutions. Furthermore, the spline boundary conditions are
chosen carefully to include the mandatory physical information on the interpolated function.
The individual spline configurations are generated by a spline Monte-Carlo already introduced in
Ref.~\cite{Brandt:2016zdy} and discussed further
in appendix~\ref{app:nI}, where we also show a set of representative examples for this interpolation.

\section{Results for the EoS}

\subsection[Thermodynamic observables at \texorpdfstring{$N_t=8$}{Nt=8}]{Thermodynamic observables at \boldmath \texorpdfstring{$N_t=8$}{Nt=8}}

\begin{figure}[t]
 \centering
\includegraphics[width=.45\textwidth]{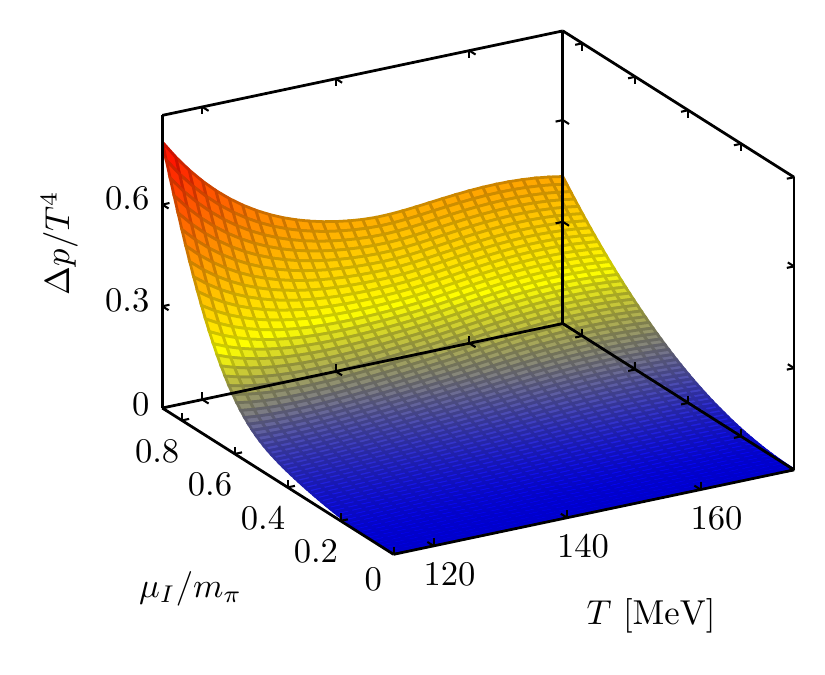}
\hspace*{0.05\textwidth}
\includegraphics[width=.45\textwidth]{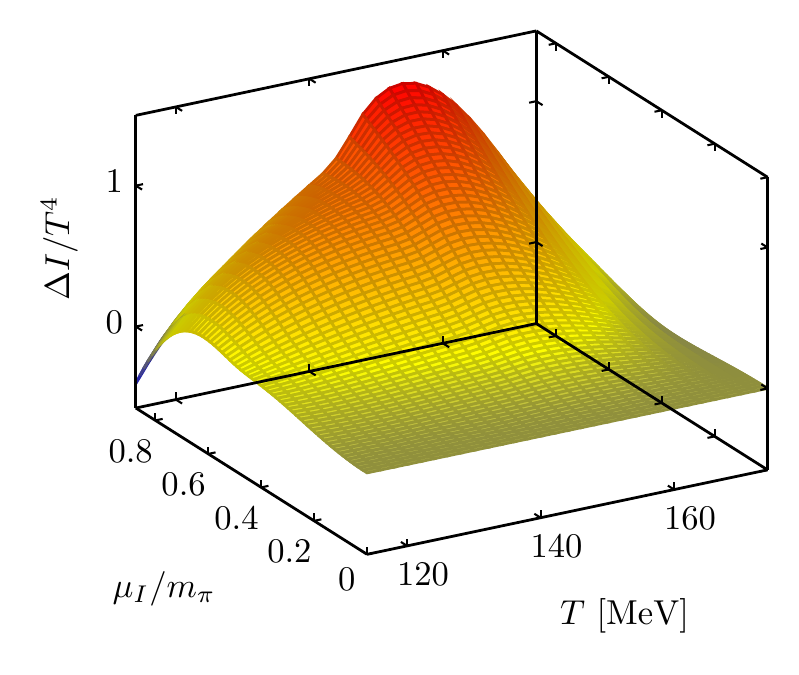} \\
\includegraphics[width=.45\textwidth]{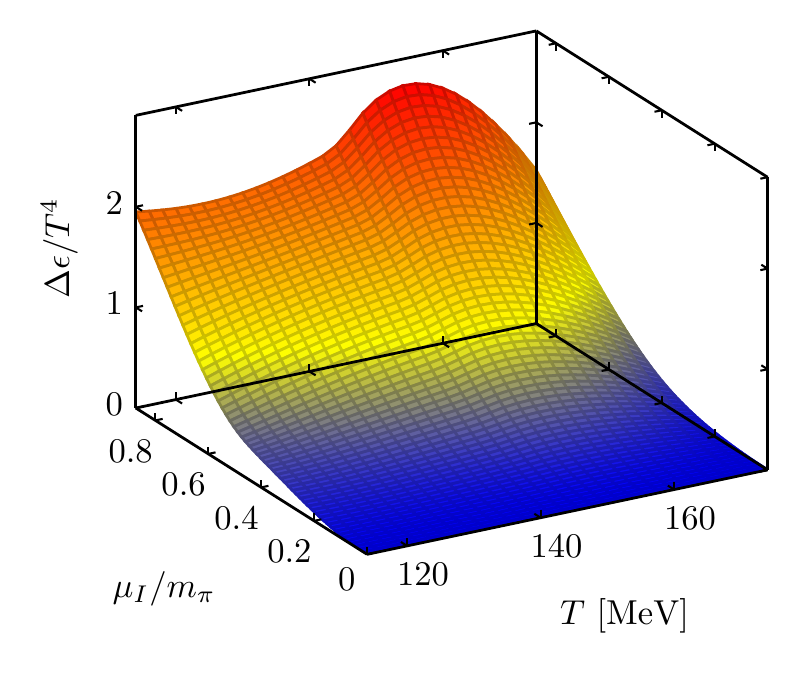}
\hspace*{0.05\textwidth}
\includegraphics[width=.45\textwidth]{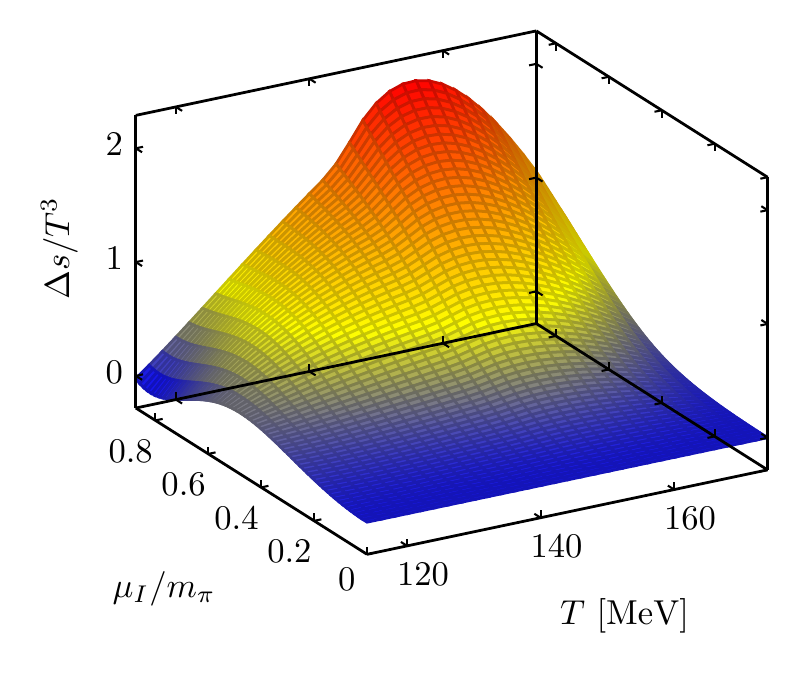}
 \caption{\label{fig:nt8-eos-dif-3d}
 Results for the modifications of the pressure (top left), the interaction measure (top right),
 the energy density (bottom left)
 and the entropy density (bottom right) due to nonzero isospin chemical potential
 on the $8\times 24^3$ lattices.
 Uncertainties are not shown for better visibility.}
\end{figure}

\begin{figure}[ht]
 \centering
\vspace*{-2mm}
\includegraphics[width=.48\textwidth]{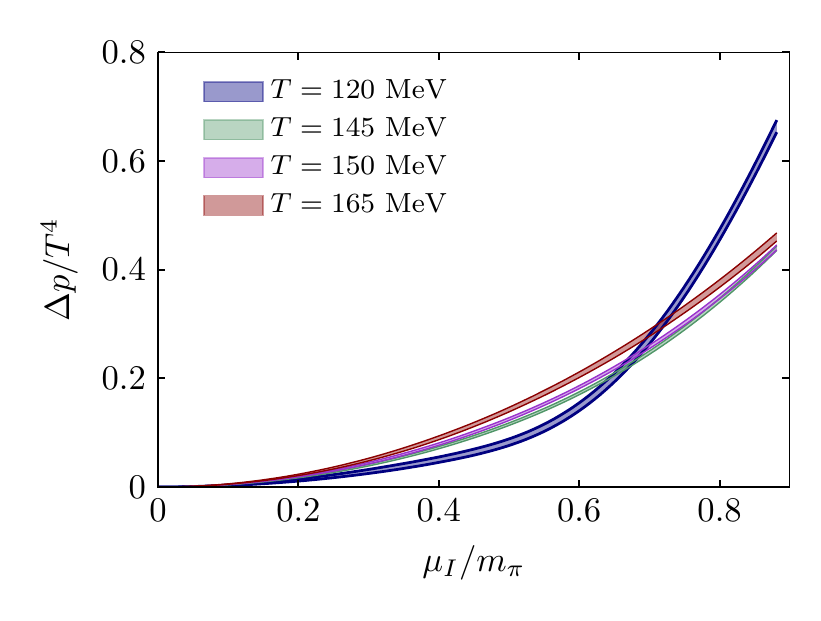}
\includegraphics[width=.48\textwidth]{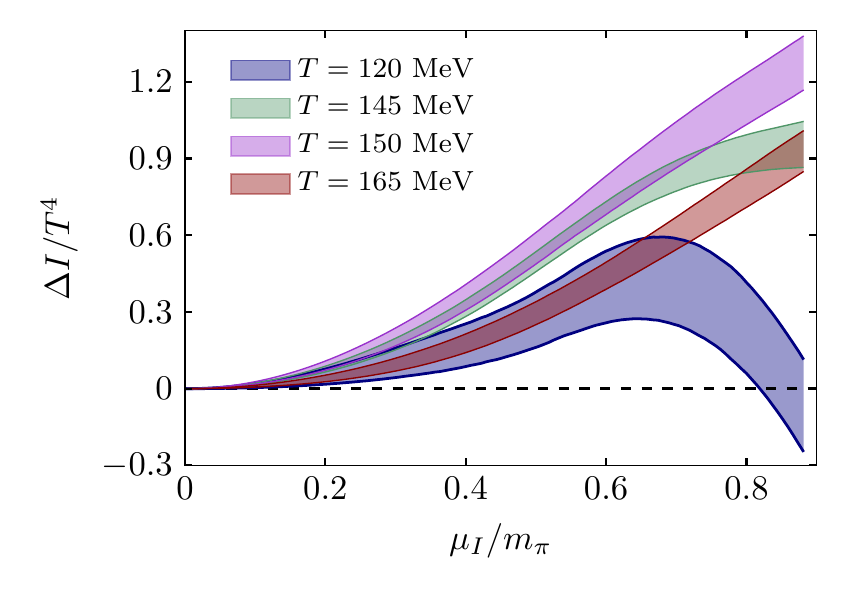}
 \caption{\label{fig:nt8-eos-dif-2d}
 Results for the modifications of the pressure (left) and the interaction measure (right)
 versus $\mui$ for different temperatures obtained on the $8\times 24^3$ lattices
 from the two-dimensional spline interpolation for $n_I(T,\mu_I)$ described in the text.}
\end{figure}

Given the two-dimensional interpolation for the isospin density $n_I(T,\mu_I)$, we can now proceed
with the computation of the
thermodynamic quantities we are interested in, namely the pressure $p$, the interaction measure $I$,
as well as energy $\epsilon$ and entropy $s$ densities using Eqs.~\cref{eq:plat}, \cref{eq:Ilat},
\cref{eq:eos-decomp} and \cref{eq:ene-ent}. The modifications of the individual observables due to
$\mu_I\neq0$ are shown in Fig.~\ref{fig:nt8-eos-dif-3d}. The pressure shows the strongest
changes due to $\mui$ for small temperatures.\footnote{Note, that this is partly also due to the
normalization by $T^4$. This normalization will become singular when we approach
the $T=0$ limit. Here we have chosen this normalization to consent to the one typically
used in the literature and to allow for easier comparisons.} 

To show the magnitude of uncertainties
and to have a more quantitative picture, we plot the pressure versus $\mui$ for different
temperatures in Fig.~\ref{fig:nt8-eos-dif-2d}. Within the BEC phase the modification of the
interaction measure shows an initial increase with $\mui$ before it decreases towards
larger $\mui$ values. This phenomenon has already been observed at
vanishing temperatures~\cite{Brandt:2018bwq,Brandt:2021yhc}, where it leads to a negative
interaction measure starting at around $\mu_I/m_\pi\approx0.84$~\cite{Brandt:2021yhc}, and
is a clear signature of the presence of the BEC in the EoS, see also Ref.~\cite{Vovchenko:2020crk}.
We will discuss the full interaction measure at $T\neq0$ below. The decrease of $\Delta I$ is
shifted to larger $\mui$ values with increasing temperature, as can be seen from the right panel
of Fig.~\ref{fig:nt8-eos-dif-2d}, where we show the interaction measure versus $\mui$ for
a few different temperatures. The decrease is no longer visible
at temperatures above around 150 MeV -- note that this is just around the edge of the BEC phase
for this particular lattice spacing. For these temperatures $\Delta I$ shows a peak in
temperature direction if $\mui/m_\pi \gtrsim 0.5$, indicating the strong influence of the
phase transition on the interaction measure. The strong change of $I$ in this
region also translates to energy and entropy densities. The former generically shows a
strong increase with $\mui$, which becomes less pronounced at larger temperatures outside of
the BEC phase. The modifications of the entropy density mainly follow those of the interaction
measure.

\begin{figure}[t]
 \centering
\vspace*{-2mm}
\includegraphics[width=.45\textwidth]{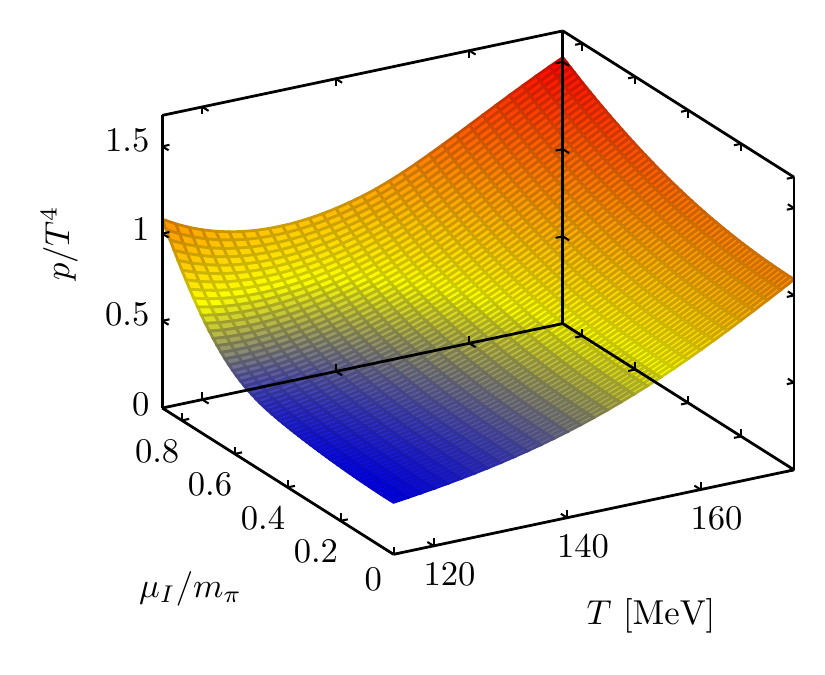}
\hspace*{0.05\textwidth}
\includegraphics[width=.45\textwidth]{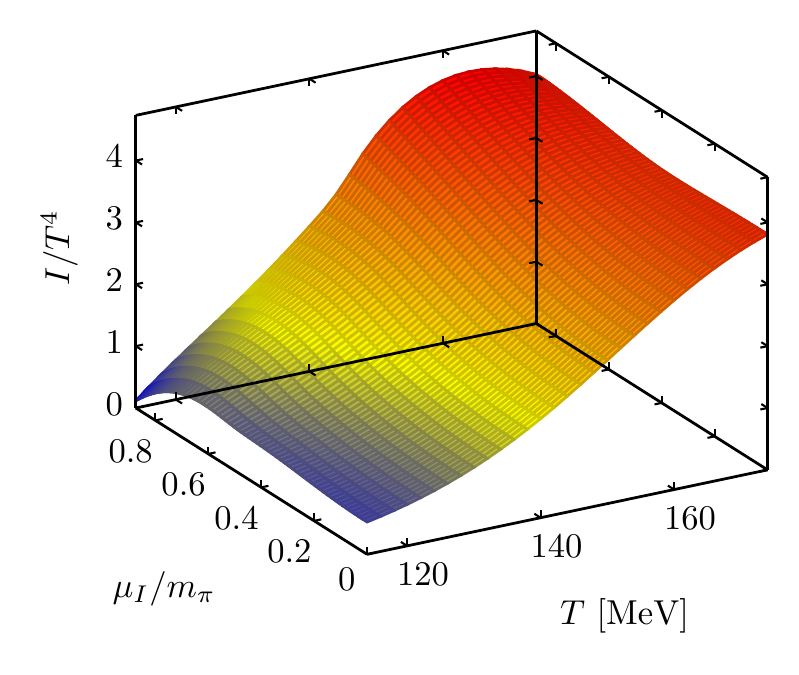} \\
\includegraphics[width=.45\textwidth]{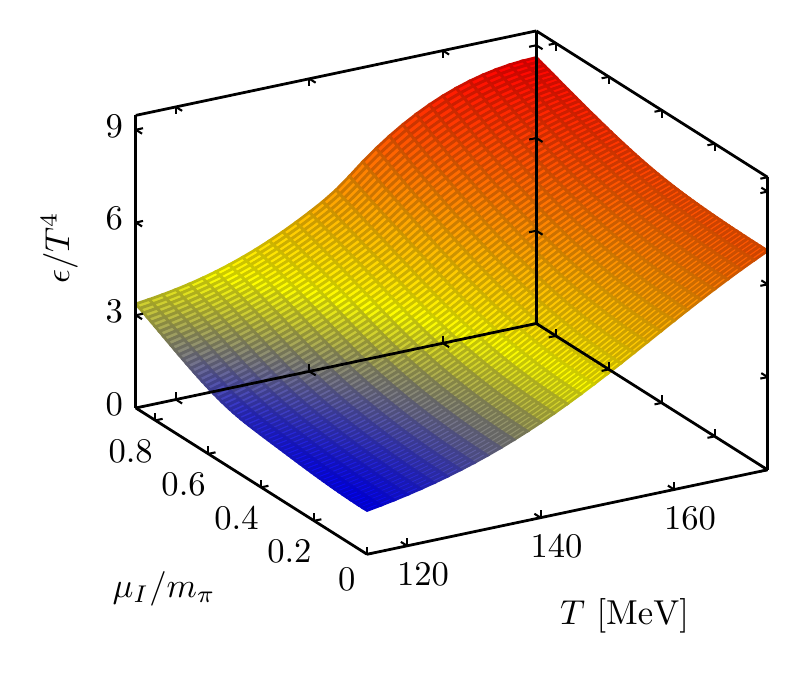}
\hspace*{0.05\textwidth}
\includegraphics[width=.45\textwidth]{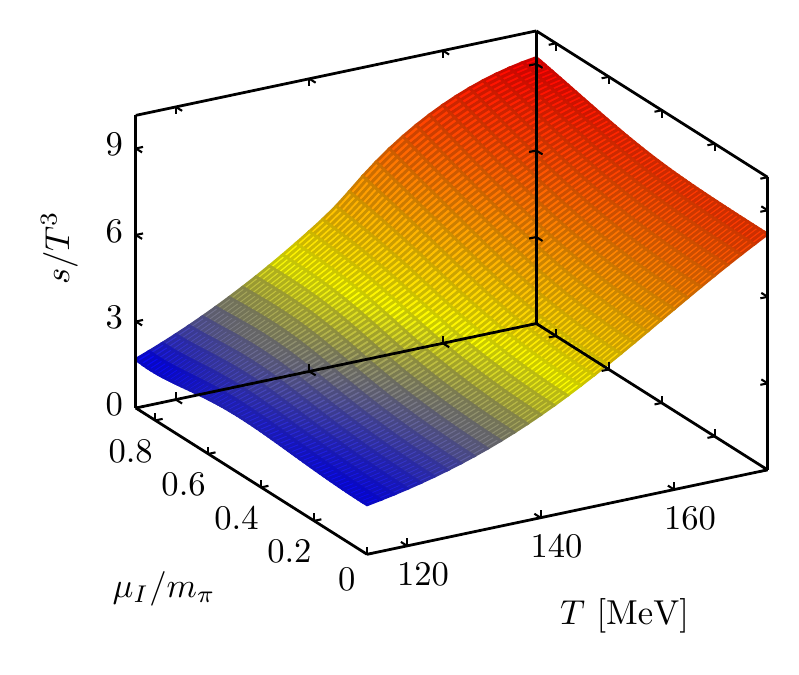}
 \caption{\label{fig:nt8-eos-3d}
 Results for the pressure (top left), the interaction measure (top right),
 the energy density (bottom left)
 and the entropy density (bottom right) in units of the temperature
 obtained on the $8\times 24^3$ lattices
 from the two-dimensional spline interpolation for $n_I(T,\mu_I)$ described in the text.
 Uncertainties are not shown for better visibility.}
\end{figure}

\begin{figure}[t]
 \centering
\vspace*{-2mm}
\includegraphics[width=.6\textwidth]{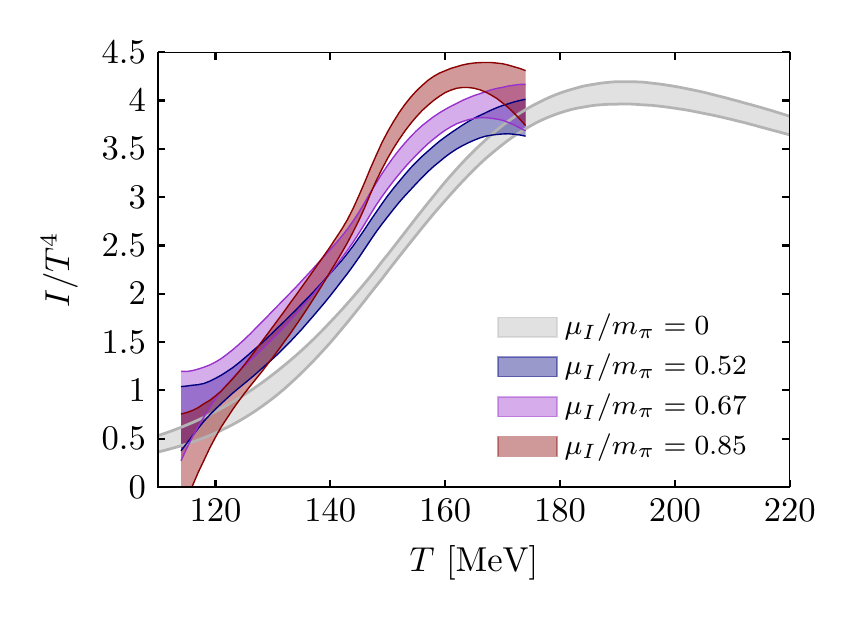}
 \caption{\label{fig:nt8-IvsT}
 Results for the interaction measure versus the temperature for different values
 of $\mui/m_\pi$ obtained on the $8\times 24^3$ lattices
 from the two-dimensional spline interpolation for $n_I(T,\mu_I)$ described in the text.
 We also compare to the $\mui=0$ results from the parameterization of
 Ref.~\cite{Borsanyi:2013bia} described in appendix~\ref{app:mui0-eos}.}
\end{figure}

To get a feeling for the overall magnitudes, we plot the full results for $p$, $I$, $\epsilon$
and $s$ in Fig.~\ref{fig:nt8-eos-3d}. To obtain the full results, we use the parameterization from
Ref.~\cite{Borsanyi:2013bia} with the coefficients introduced in appendix~\ref{app:mui0-eos}
for the $\mui=0$ quantities from Eq.~\cref{eq:eos-decomp}.\footnote{Note that this procedure uses continuum results at $\mui=0$,
but our results for the modifications by $\mui$ obtained at
non-vanishing lattice spacing.} We observe that the pressure generically rises with $T$
and $\mui$. The exception is again the small $T$ and large $\mui$ region, where the increase seen
in the plot, however, stems from the normalization with the temperature as explained above.
A similar monotonic rise with $T$ and $\mui$ is generically observed for the energy density,
even though the increase tends to become less pronounced for larger temperatures outside of
the BEC phase. Note that the entropy density vanishes in the zero temperature limit for
all values of $\mu_I$.

For the interaction measure we can clearly observe the decrease with $\mui$
deep in the BEC phase for small temperatures. Furthermore, we also see a flattening and the
onset of a peak-like
structure for temperatures around the BEC phase boundary. This is an interesting observation.
A peak of the interaction
measure is present at $\mui=0$~\cite{Borsanyi:2013bia}, see the gray curve in Fig.~\ref{fig:nt8-IvsT}, at a temperature
of around 190 MeV, i.e., above the phase transition. The observed flattening of $I$ might thus
be the onset of this peak structure, shifted towards smaller temperatures for increasing $\mui$. If this is the case, the peak approaches the thermal crossover at smaller values
of $\mui$, until it becomes mostly consistent with the boundary of the BEC phase, see Fig.~\ref{fig:phd}, around
$\mui/m_\pi\approx0.8$. This is further visualized in Fig.~\ref{fig:nt8-IvsT}, where we show the
interaction measure versus the temperature for different $\mui/m_\pi$. Together with the shift
of the onset of the plateau, we also observe the development of a narrower peak structure
for larger $\mu_I$ values. This might be
an effect of the second order phase transition at the BEC phase boundary. A similar decrease of
the temperature of the maximum with the chemical potential is also observed at nonzero baryon
chemical potential, see~\cite{Borsanyi:2012cr,Guenther:2017hnx} (as well as the large modifications
of the interaction measure around $T\approx155$~MeV observed in Ref.~\cite{Bollweg:2022fqq}), for instance.

\subsection[Lattice artifacts at \texorpdfstring{$N_t=8$}{Nt=8}, 10 and 12]{\boldmath Lattice artifacts at \texorpdfstring{$N_t=8$}{Nt=8}, 10 and 12}

\begin{figure}[t]
 \centering
\vspace*{-2mm}
\includegraphics[width=.48\textwidth]{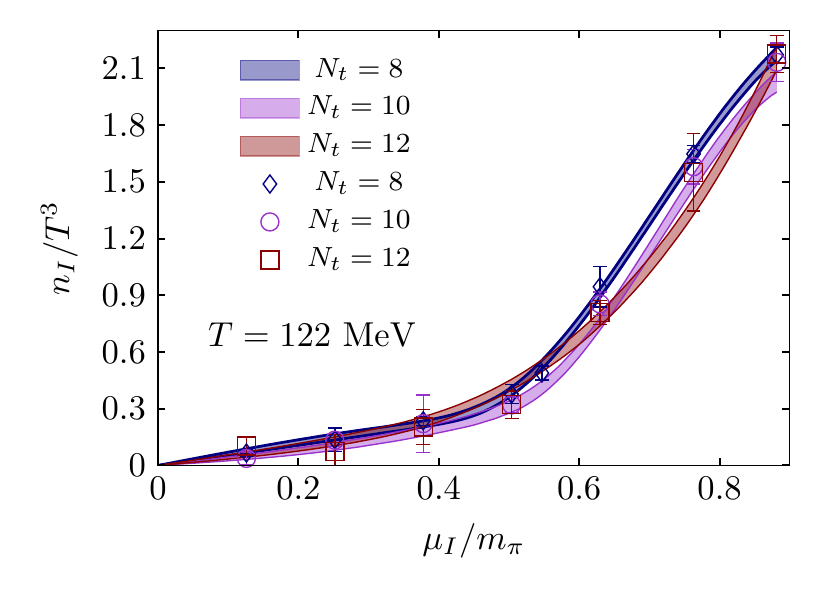}
\includegraphics[width=.48\textwidth]{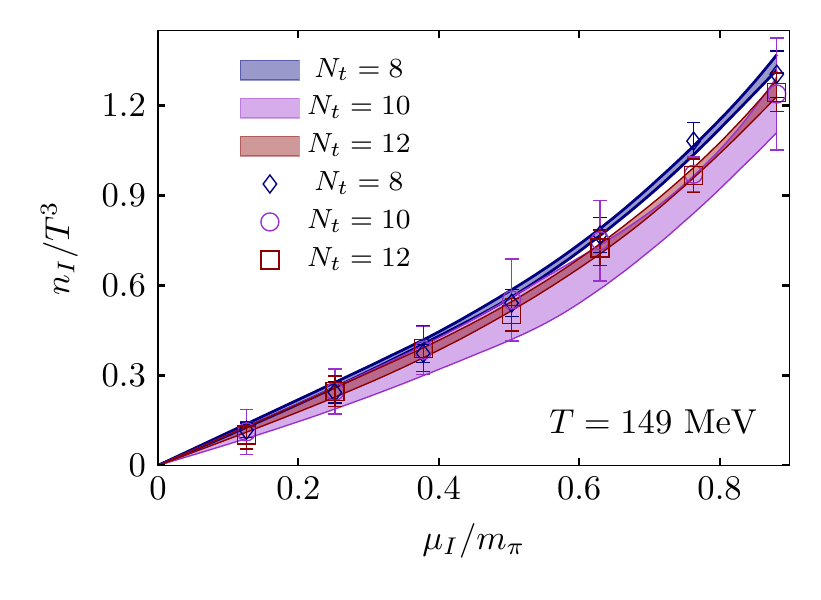}
 \caption{\label{fig:nI-latart}
 Results for the isospin density obtained on $8\times24^3$, $10\times28^3$ and $12\times36^3$ lattices, together with the spline interpolation ($T=122$ and $149$ MeV slices of the global two-dimensional spline fit).}
\end{figure}

So far we have discussed results obtained for one value of $N_t$, corresponding to one
particular lattice spacing for each value of $T$. To obtain continuum results we have
to increase $N_t$ while keeping the physical temperature and the aspect ratio $N_s/N_t$ fixed.
Unfortunately, for the isospin density a well controlled
continuum extrapolation is not possible with the $N_t=8,\,10$ and 12 lattices currently at
our disposal. To show this, we plot the results for $n_I$ obtained from
$8\times24^3$, $10\times28^3$ and $12\times36^3$ lattices for two different temperatures
in Fig.~\ref{fig:nI-latart}. For both temperatures, we observe that at small $\mui$ the results
of the $N_t=12$ lattices lie between those of the $N_t=8$ and 10 lattices. This indicates that
higher order lattice artifacts are still present in this region and impede a proper
continuum extrapolation. This behavior is in agreement with the large lattice artifacts
observed for the leading order Taylor expansion coefficient in the direction of the isospin
chemical potential, see, e.g.\ Ref.~\cite{Borsanyi:2011sw}.
The ordering of the results from the different $N_t$ values remains even for larger
values of $\mui$ at $T=149$~MeV. At smaller temperatures the data rearranges in the vicinity
of the phase transition to the BEC phase, such that the magnitude of $n_I$ increases from
$N_t=12$ to $N_t=10$ and 8.

\begin{figure}[t]
 \centering
\vspace*{-2mm}
\includegraphics[width=.45\textwidth]{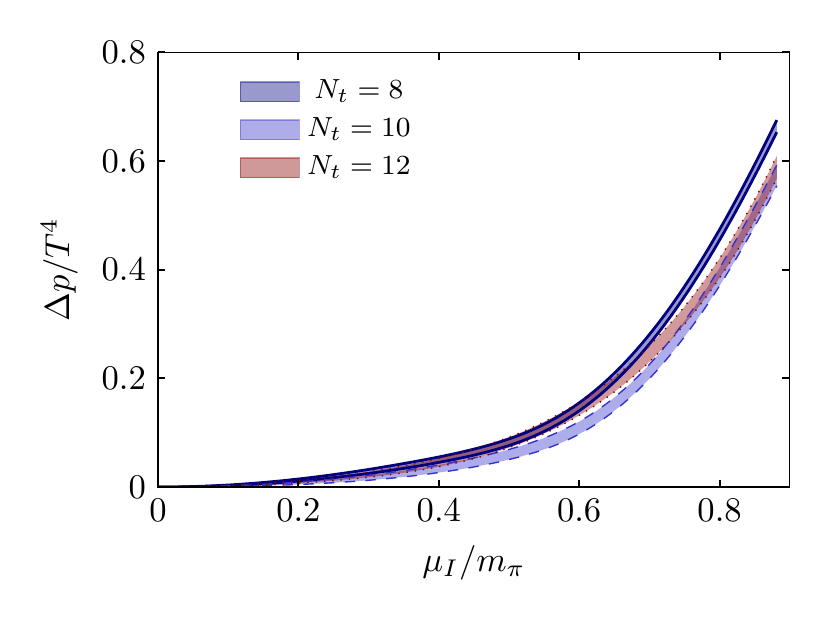}
\hspace*{0.05\textwidth}
\includegraphics[width=.45\textwidth]{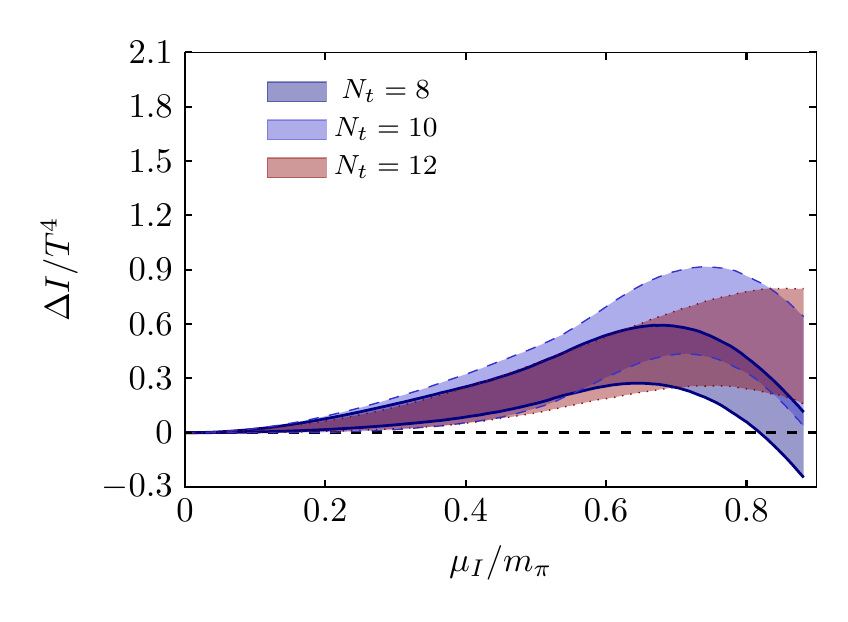} \\
\includegraphics[width=.45\textwidth]{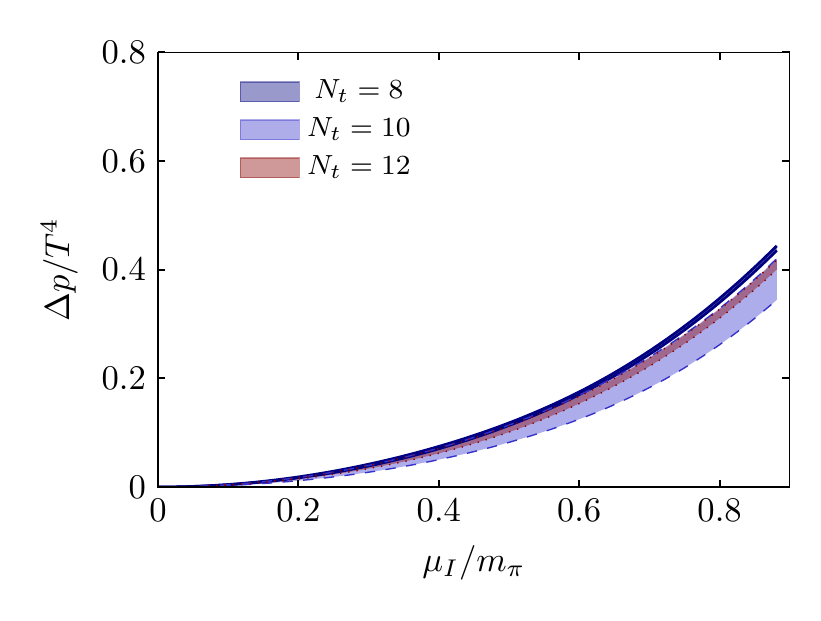}
\hspace*{0.05\textwidth}
\includegraphics[width=.45\textwidth]{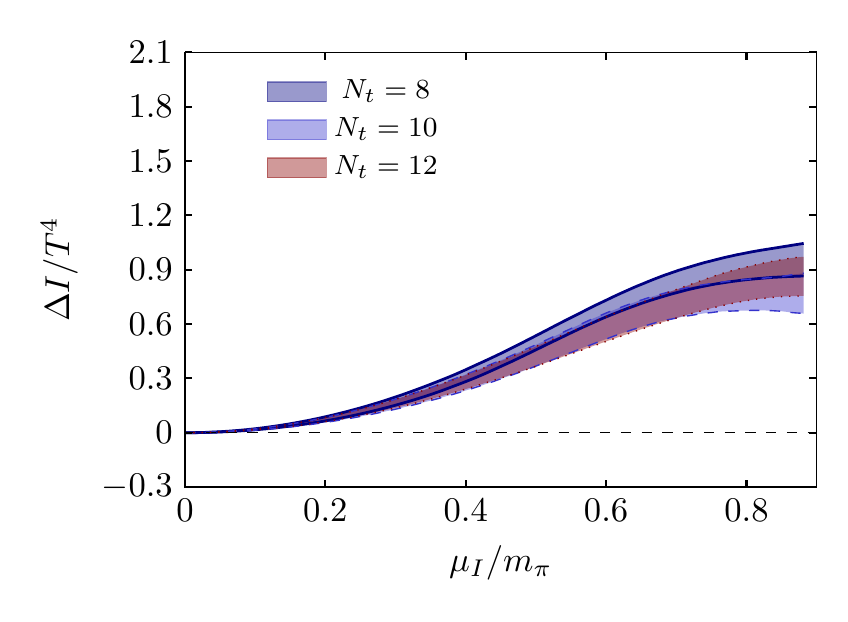} \\
\includegraphics[width=.45\textwidth]{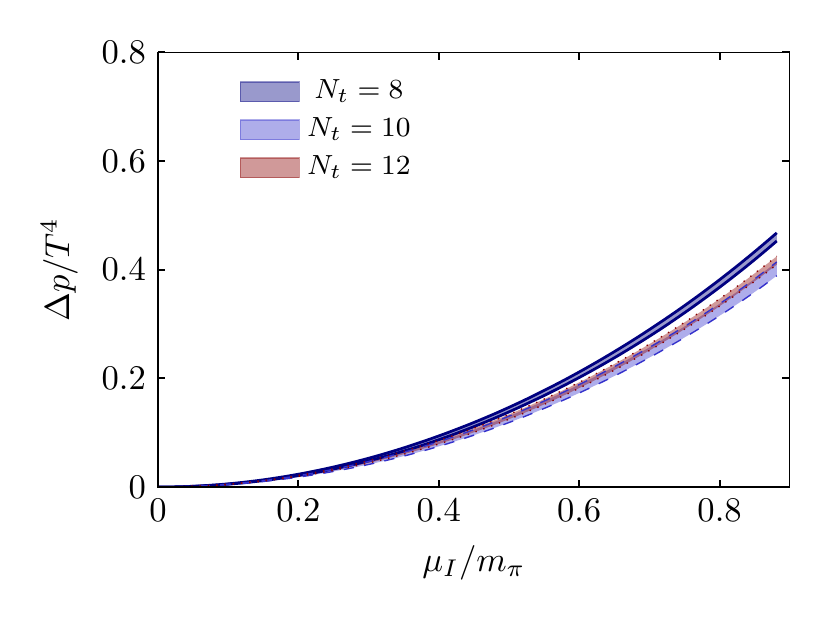}
\hspace*{0.05\textwidth}
\includegraphics[width=.45\textwidth]{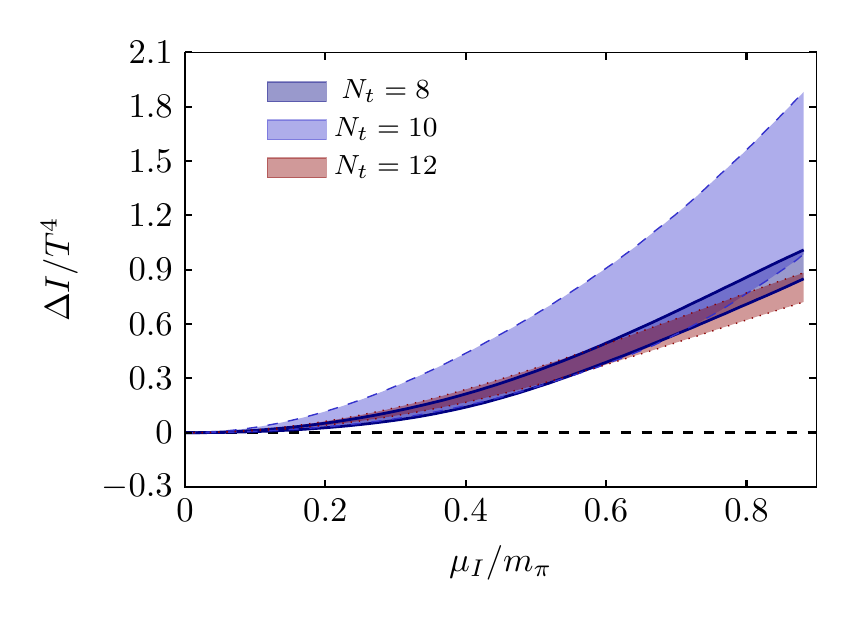}
 \caption{\label{fig:eos-temps}
 Results for the pressure (left) and the interaction measure (right)
 for temperatures $T=120$, 145 and 165~MeV (top to bottom) from
 $N_t=8$, 10 and 12 lattices.}
\end{figure}

While a direct continuum extrapolation for the isospin density is not possible with the
current dataset, we can still look at the magnitude of lattice artifacts for the other
observables related to the EoS. We show the results for the pressure and the interaction
measure from the lattices at different values of $N_t$ for three different temperatures
in Fig.~\ref{fig:eos-temps}. The plot shows that the pressure, as the direct integral over
isospin density, suffers from similar lattice artifacts as $n_I$ and again the results
from the $N_t=12$ lattice are located between those of the $N_t=8$ and 10 lattices.
For the interaction measure the situation is a bit different in the sense that all
results typically overlap within the (comparably large) uncertainties.

\subsection[The phase diagram in the \texorpdfstring{$T$-$n_I$}{T-nI} plane]{\boldmath The phase diagram in the \texorpdfstring{$T$-$n_I$}{T-nI} plane}
\label{sec:pd-ni-T}

\begin{figure}[t]
 \centering
\vspace*{-2mm}
\includegraphics[width=.48\textwidth]{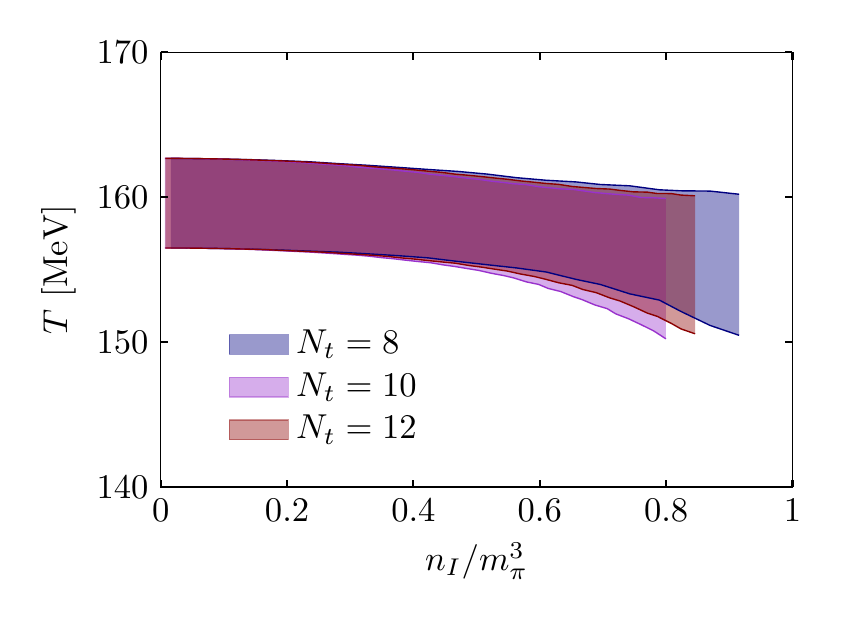}
\includegraphics[width=.48\textwidth]{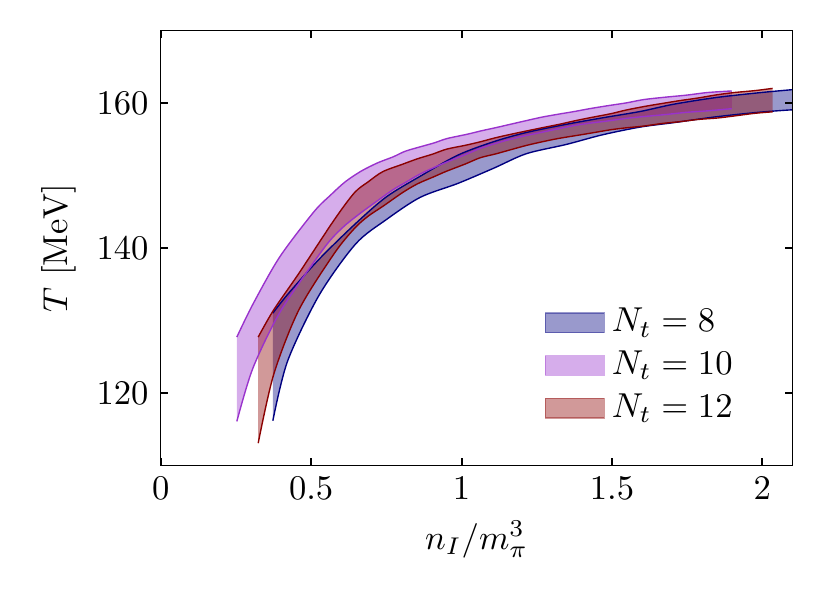}
 \caption{\label{fig:nI-ph-bounds}
 Results for the chiral crossover temperature (left) and the BEC phase boundary (right) in the $T$-$n_I$ plane for
 different values of $N_t$, based on our results from Ref.~\cite{Brandt:2017oyy}.}
\end{figure}

From the interpolation of $n_I$ we can also extract the phase diagram in the $T$-$n_I$ plane, of
which a preliminary version has been presented in Ref.~\cite{Brandt:2021yhc}.
To this end we use the phase boundaries in the $T$-$\mu_I$ plane from Ref.~\cite{Brandt:2017oyy}
for the individual temporal extents, $N_t=8$ to 12, and determine the value of $n_I$ on
these phase boundaries. The results for the chiral crossover and the BEC phase boundary for all
$N_t$ are shown in Fig.~\ref{fig:nI-ph-bounds}. While the chiral crossover does not show any lattice spacing dependence, the BEC phase boundary exhibits slight lattice artifacts.

\begin{figure}[t]
 \centering
\vspace*{-2mm}
\includegraphics[width=.48\textwidth]{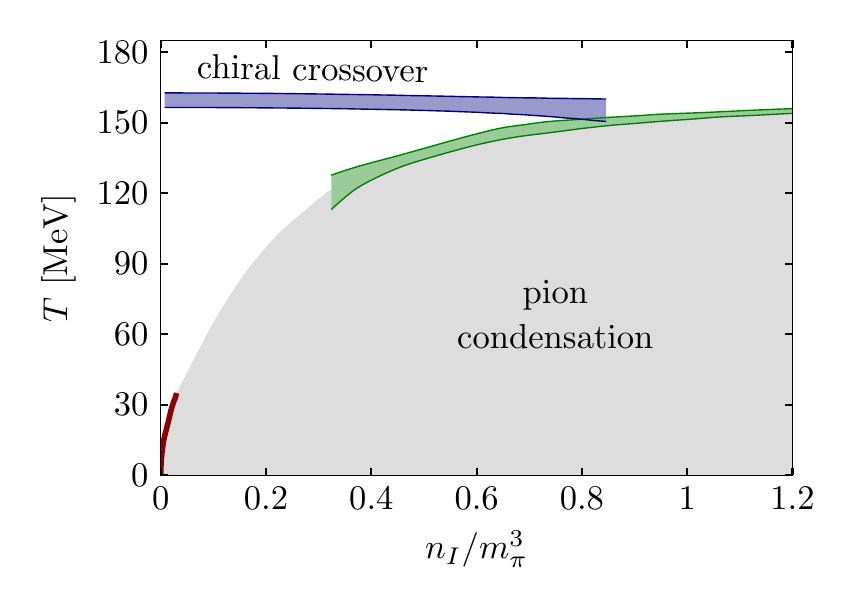}
 \caption{\label{fig:nI-ph-diag}
 Results for the phase diagram in the $T$-$n_I$ plane for $N_t=12$ together with the results from
 next-to-leading order chiral perturbation theory of Ref.~\cite{Adhikari:2020kdn} (red curve).
 The continuation of the $N_t=12$ pion condensation boundary to the chiral perturbation theory
 line merely serves to guide the eye.}
\end{figure}

Due to the Silver blaze property at $T=0$, the BEC phase transition starts at $n_I=T=0$.
The BEC phase boundary 
at small temperatures and isospin denstities has been calculated via next-to-leading order
chiral perturbation theory~\cite{Adhikari:2020kdn}. We show those results up to $T=30$~MeV (where the phase boundary in chiral perturbation theory starts to deviate significantly from the lattice result)
together with the
phase boundaries from the $N_t=12$ lattices in Fig.~\ref{fig:nI-ph-diag}.

\section{The speed of sound}
\label{sec:cs-comp}

\subsection{Computation from the interpolation of the isospin density}

Another interesting observable related to the EoS is the speed of sound $c_s$.
The isentropic speed of sound, i.e., the speed of a sound wave travelling isentropically
through the plasma, is defined as
\be
\label{eq:cs-def}
c_s^2 = \left. \frac{\partial p}{\partial \epsilon} \right|_{\rm iso} = 
\frac{\partial_\xi\, p}{\partial_\xi\, \epsilon} \,,
\ee
where the subscript ${\rm iso}$ refers to the derivative taken in the direction of
isentropic trajectories in phase space, i.e., for QCD at generic nonzero quark chemical
potentials in the direction where
\be
\label{eq:cs-iso-cond}
\frac{s}{n_q}={\rm const} \quad \forall\:\text{flavours}\:q
\qquad(q=u,\,d,\,s \:\:\text{for}\:\: N_f=2+1) \,,
\ee
and we have introduced the directional derivative $\partial_\xi$ in this direction in parameter space.
For a pure isospin chemical potential the only relevant density is the isospin density,
so that the condition~\cref{eq:cs-iso-cond} reduces to
\be
\label{eq:cs-iso-iso}
\frac{s}{n_I}={\rm const} \,, \quad \text{or} \quad \partial_\xi \left(\frac{s}{n_I}\right) = 0 \,,
\ee
and we can write the directional derivative as
\be
\label{eq:cs-dir-der}
\partial_\xi = \vec{\xi} \cdot \left(\begin{array}{c} \partial_T \\ \partial_{\mui} \end{array}\right)
= \cos\alpha \,\partial_T + \sin\alpha \,\partial_{\mui} \,.
\ee

Combining Eqs.~\cref{eq:cs-iso-iso} and~\cref{eq:cs-dir-der} we can compute the angle $\alpha$ via
\be
\tan\alpha = - \left(n_I \frac{\partial s}{\partial T} - s \frac{\partial n_I}{\partial T}
\right)\:\left(n_I \frac{\partial s}{\partial \mui} - s \frac{\partial n_I}{\partial \mui}
\right)^{-1}\,,
\ee
where all quantities and derivatives can be obtained analytically from the spline interpolation
for $n_I$ and the analytic form for the interaction measure at $\mui=0$ from appendix~\ref{app:mui0-eos}.
Once $\alpha$ has been obtained, one can similarly analytically compute the directional
derivatives of $p$ and $\epsilon$ in Eq.~\cref{eq:cs-def}.

Another quantity of interest related to the speed of sound, in particular for astrophysical
and cosmological applications, is the polytropic index (see~\cite{Annala:2019puf}, as well as the lectures~\cite{Silbar:2003wm,Sagert:2005fw})
\be
\gamma = \left. \frac{d\log p}{d\log\epsilon} \right|_{\rm iso} = \frac{\epsilon}{p} \,c_s^2 \,.
\ee
In the conformal limit, approached by QCD at asymptotically large densities or temperatures,
it takes a value of $\gamma=1$, while in the hadronic regime conformal symmetry is broken due
to spontaneous chiral symmetry breaking, leading to large values of $\gamma$ in the range
of $\gamma\gtrsim 2$ (see the discussion in Ref.~\cite{Annala:2019puf}).
Consequently, $\gamma$ can be seen as a measure in the EoS to distinguish between regions of
hadron dominated matter (confined) or matter dominated by free quarks (quarkyonic/deconfined).
In the study of Ref.~\cite{Annala:2019puf} a value of $\gamma=1.75$ has been introduced
to distinguish between these two types of matter in neutron star cores. Finally, we also
look at the normalized trace anomaly~\cite{Fujimoto:2022ohj},
\be
 \Delta = \frac{1}{3} - \frac{p}{\epsilon} = \frac{I}{3\,\epsilon} \,,
\ee
which should be a number between $-2/3$ and $1/3$ due to causality and thermodynamic stability.
Furthermore, in Ref.~\cite{Fujimoto:2022ohj} it has been argued that $\Delta\geq0$.

\subsection{Speed of sound at vanishing temperature}
\label{sec:cs-t0dis}

Before discussing the results for the isentropic speed of sound in the parameter space of
nonzero $(T,\,\mui)$, it is instructive to look at the limiting case of vanishing temperature.
An initial study of the EoS at $T=0$ on a coarse lattice with
$a\approx 0.29$~fm has already been presented in
Refs.~\cite{Brandt:2018bwq,Brandt:2021yhc}.
Here we will present new results for the speed of sound at $T=0$,
obtained on $24^3\times32$ and $32^3\times48$ lattices at lattice spacings of $a\approx0.22$~fm
and  $a\approx0.15$~fm, respectively, including
data up to $\mui/m_\pi\approx1$. The results for these lattice spacings have already been presented
partly in Ref.~\cite{Brandt:2022fij} where they also have been compared to the $a\approx 0.29$~fm data.

\begin{figure}[t]
 \centering
\vspace*{-2mm}
\includegraphics[width=.48\textwidth]{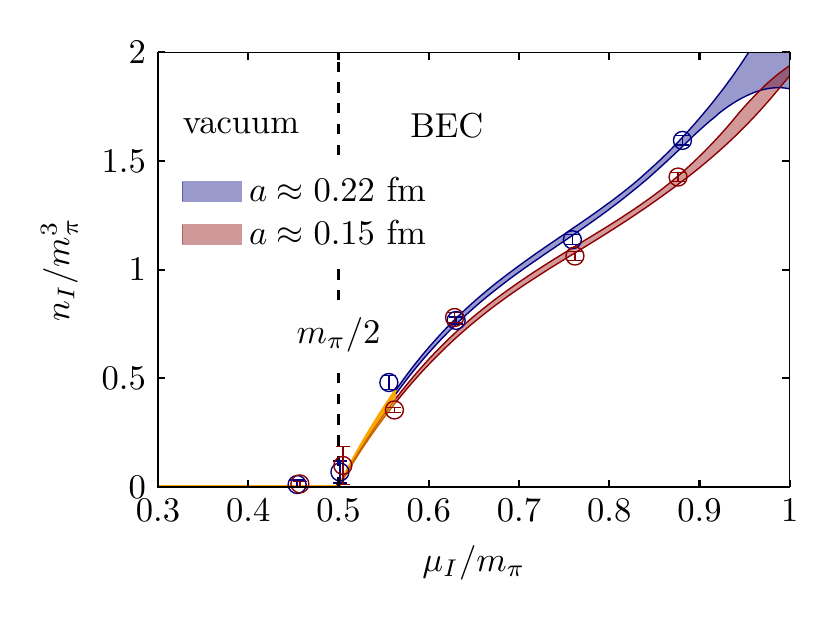}
\includegraphics[width=.48\textwidth]{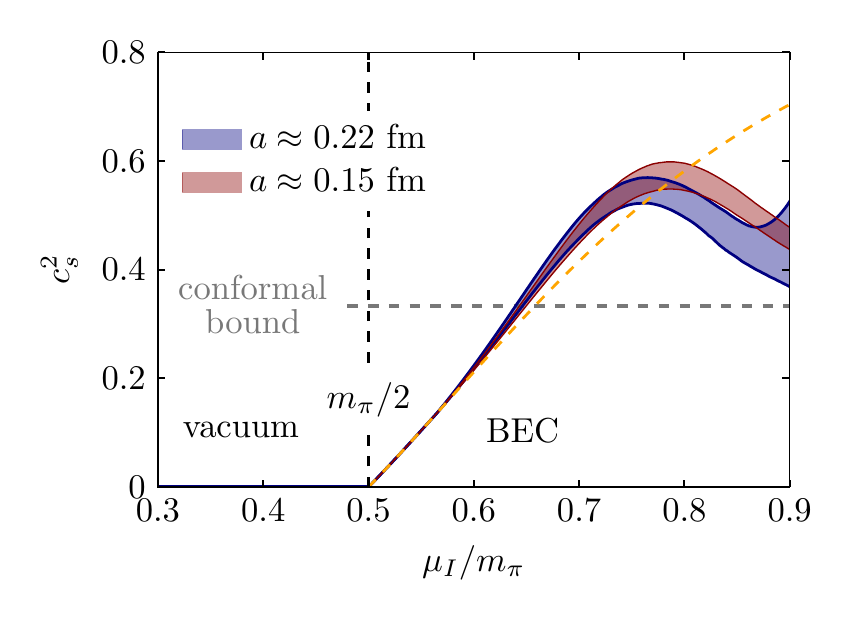}
 \caption{\label{fig:cs-T0}
 {\bf Left:} Results for the isospin density together with the spline interpolation at $T=0$
 obtained on $24^3\times32$ and $32^3\times48$ lattices and lattice spacings
 of $a\approx0.22$ and 0.15~fm. The yellow
 part of the curve is obtained directly from the chiral perturbation theory expression
 for $n_I$.
 {\bf Right:} Results for the isentropic speed of sound at $T=0$, obtained from the spline
 interpolation of the left panel. Also shown are the chiral perturbation theory result
 (dashed yellow line) with the pion decay constant obtained from the fit discussed in
 the text as well as the conformal bound~\cite{Cherman:2009tw} (dashed gray line).}
\end{figure}

The starting point for the extraction of the EoS at zero temperature is again the isospin
density, from which one can obtain the pressure and, consequently, all other thermodynamic
quantities, using Eq.~\cref{eq:plat}. Due to the Silver Blaze property, the isospin density
vanishes outside of the BEC phase at $T=0$. In practice, the simulations are performed at
a small but non-vanishing temperature, so that residual temperature effects on $n_I$ need to
be corrected in the vicinity of the transition. As already done in Ref.~\cite{Brandt:2018bwq}
we use chiral perturbation theory~\cite{Son:2000xc} to correct for these $T\neq0$ effects.
In particular, we fit the results for $n_I$ for the two smallest values of $\mui$ within
the BEC phase, i.e., we include the data points up to $\mui/m_\pi\approx0.65$, to the
chiral perturbation theory expression for $n_I$
(see Eq.~(B1) in appendix~B of Ref.~\cite{Brandt:2018bwq}). For our present ensembles this fit yields
values of $f_\pi=136(2)$ and $130(3)$~MeV, respectively, in good agreement with the physical
value and the result obtained from the fit in Ref.~\cite{Brandt:2018bwq} for the lattice
with $a\approx 0.29$~fm.
To obtain a smooth interpolation for the isospin density, we again perform a model independent
spline interpolation of the remaining data points via a spline Monte-Carlo as discussed above,
where all splines are matched to chiral perturbation theory.\footnote{For a smooth matching up
to the second $\mui$-derivative, relevant for $c_s$, the matching of the spline is done at half
the distance between the second data point in the BEC phase and the BEC phase boundary and
we have included 20 additional data points generated from chiral perturbation theory, equally
separated in the remaining interval up to the second data point, in the fit. The latter is
relevant for a smooth matching of the second derivative.} The resulting interpolation
of the isospin density is shown in the left panel of Fig.~\ref{fig:cs-T0}.

At $T=0$ the condition of Eq.~\cref{eq:cs-iso-iso} is trivially fulfilled since $s$ vanishes.
Thus, the directional derivative is equivalent to the $\mui$-derviative,
$\partial_\xi=\partial_{\mui}$. The resulting derivatives of $p$ and $\epsilon$ in
Eq.~\cref{eq:cs-def} can again be computed analytically. The results for the square of the
isentropic speed of sound are shown in the right panel of Fig.~\ref{fig:cs-T0}, together
with the conformal bound~\cite{Cherman:2009tw} as a gray dashed line.
We observe that the squared speed of sound crosses the conformal bound at
$\mui/m_\pi\approx0.643(4)$ and $\mui/m_\pi\approx0.646(5)$, for $a\approx0.22$ and 0.15~fm,
respectively, and reaches a peak at
\be
\label{eq:cs-peak}
\left. \frac{\mui}{m_\pi} \right|_{{\rm max}(c_s)} = \left\{\begin{array}{ll} 0.76(4) & \quad \text{with} \quad {\rm max}(c_s^2) = 0.55(3) \quad \text{for} \quad a\approx0.22 \:\text{fm} \vspace*{2mm} \\ 0.78(5) & \quad \text{with} \quad {\rm max}(c_s^2) = 0.57(3) \quad \text{for} \quad a\approx0.15 \:\text{fm} \,. \end{array} \right.
\ee
We note that our finding of $c_s^2>1/3$, as well as the development of a peak is in good
agreement with recent results obtained in two-color QCD~\cite{Iida:2022hyy,Itou:2022ebw}
(see also~\cite{Kojo:2021hqh}). Furthermore,
similar peaks in the speed of sound appear in quarkyonic 
models~\cite{McLerran:2018hbz,Jeong:2019lhv,Kovensky:2020xif,Kojo:2021ugu}.
At larger $\mui$, the speed of sound decreases and, on general grounds and according to
perturbation theory~\cite{Annala:2019puf},
is expected to approach the conformal bound asymptotically from below. For our values of
$\mui$, we currently do not see the decrease below the conformal bound.
This would require simulations at yet higher isospin chemical potentials.

\begin{figure}[t]
 \centering
\vspace*{-2mm}
\includegraphics[width=.48\textwidth]{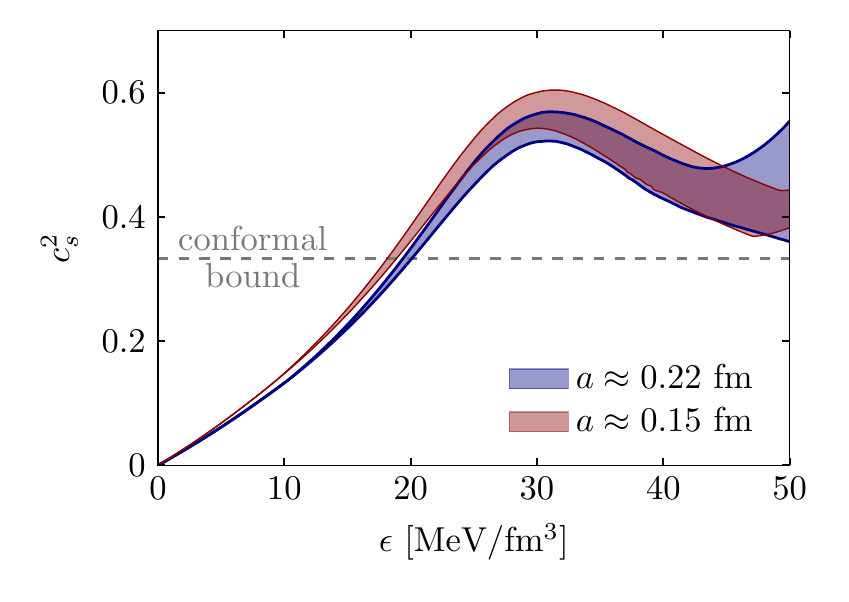}
\includegraphics[width=.48\textwidth]{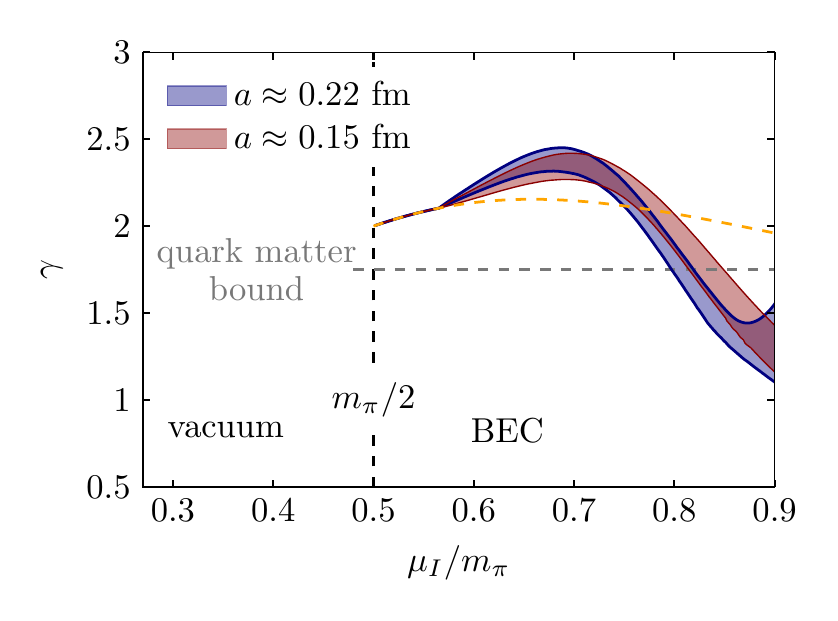}
\includegraphics[width=.48\textwidth]{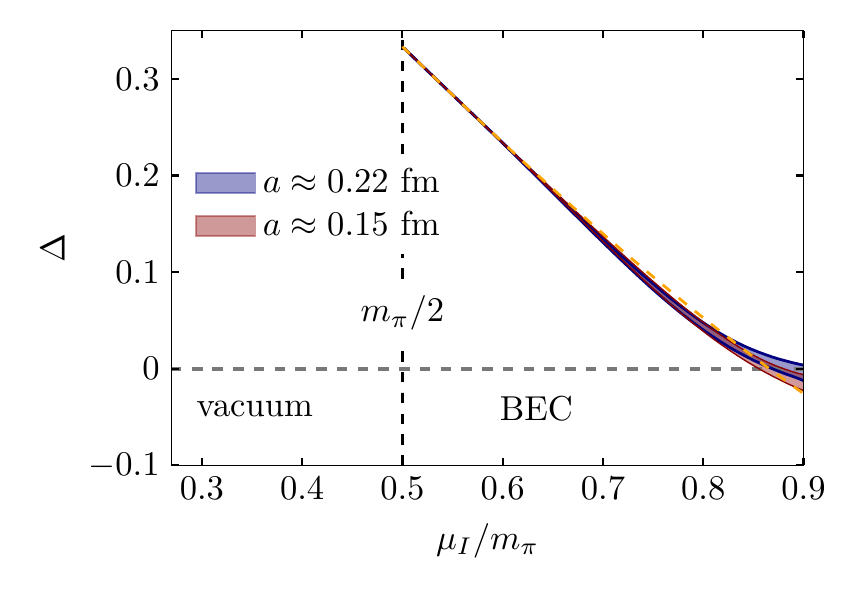}
 \caption{\label{fig:cs-T0-gam}
 {\bf Left:} Results for the isentropic speed of sound at $T=0$ versus the energy density
 in the BEC phase. Also shown is the conformal bound~\cite{Cherman:2009tw} (gray dashed line).
 {\bf Right:} Results for the polytropic index $\gamma$ at $T=0$. Also shown is the result from chiral perturbation theory (yellow dashed line) and the ``quark matter bound'' introduced in Ref.~\cite{Annala:2019puf} (gray dashed line).
 {\bf Bottom:} Results for the normalized conformal anomaly $\Delta$ at $T=0$,
 together with the result from chiral perturbation theory (yellow dashed line).}
\end{figure}

To allow for contact with recent studies on the EoS in neutron
stars~\cite{Annala:2019puf,Somasundaram:2021clp,Annala:2021gom,Altiparmak:2022bke,Ecker:2022xxj,Marczenko:2022jhl},
we plot $c_s^2$ versus the energy density in the upper left panel of Fig.~\ref{fig:cs-T0-gam}.
Comparing to typical energy densities reached in the most massive neutron star cores
(which are of the order of $10^3\text{ MeV}/\text{fm}^3$, see e.g.\ Refs.~\cite{Somasundaram:2021clp,Marczenko:2022jhl}),
we see that the speed of sound reaches values which are larger than the conformal bound
already for around one to two orders of magnitude smaller energy densities. In the upper right panel
of Fig.~\ref{fig:cs-T0-gam}, we show the polytropic index versus $\mui$. Just at
the onset of pion condensation, $\gamma$ assumes a value of $\gamma\geq2$,
greater than the ``quark matter bound'' introduced in Ref.~\cite{Annala:2019puf}, in agreement with the
prediction from chiral perturbation theory (yellow dashed curve). It then increases
with $\mui$ until it reaches its maximal value $\gamma\approx2.4$ around $\mui/m_\pi$
between 0.67 and 0.72, depending on the lattice spacing.
Further increasing $\mui$, the polytropic index decreases below 1.75 and is seen to approach its conformal
value of $\gamma=1$ asymptotically. We note that the crossing of the ``quark matter bound'' might provide an alternative definition for the BEC-BCS crossover~\cite{Brandt:2019hel,Cuteri:2021hiq}, where effective degrees of freedom change from pions to Cooper pairs of $u$ and $\bar d$ quarks.

Finally, the normalized trace anomaly $\Delta$ is plotted in the
bottom panel of Fig.~\ref{fig:cs-T0-gam}. It starts at $1/3$ at the onset of the BEC phase,
in good agreement with chiral perturbation theory, and decreases towards larger $\mu_I$.
Eventually it becomes negative between $\mui/m_\pi$ of 0.85 to 0.9 on the border of our
parameter interval. Confirming the prediction of chiral perturbation theory, this shows a specific counter-example to the claim that $\Delta$ would be strictly positive in QCD.

\subsection{Speed of sound at nonzero temperatures}

\begin{figure}[t]
 \centering
\vspace*{-2mm}
\includegraphics[width=.45\textwidth]{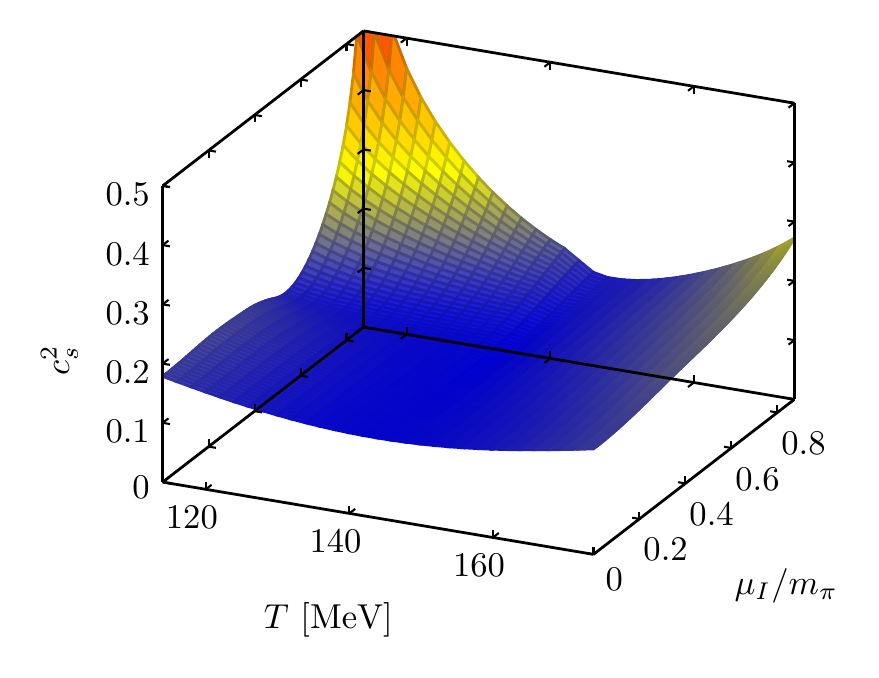}
\hspace*{0.05\textwidth}
\includegraphics[width=.45\textwidth]{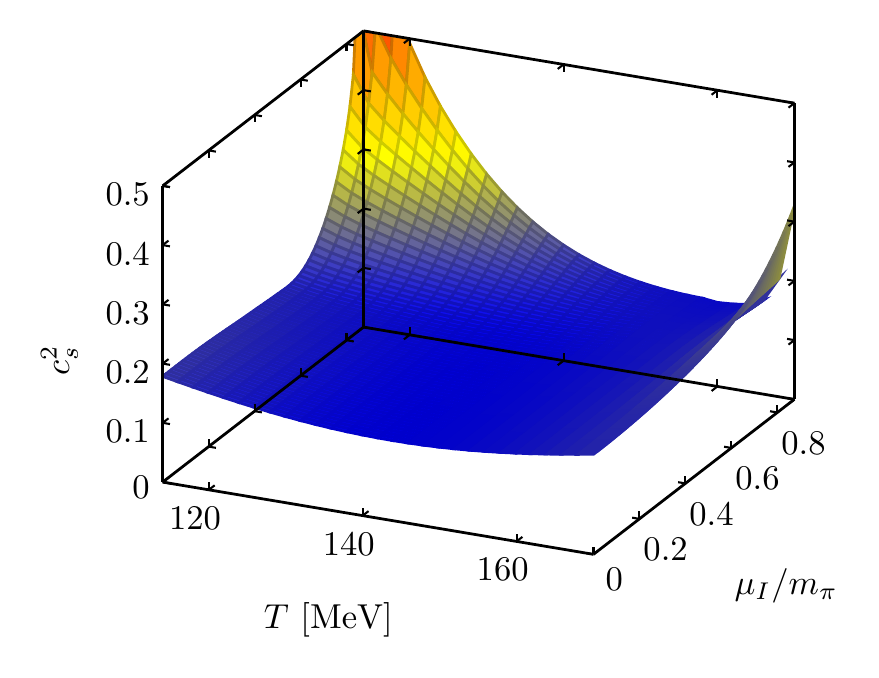} \\
\includegraphics[width=.45\textwidth]{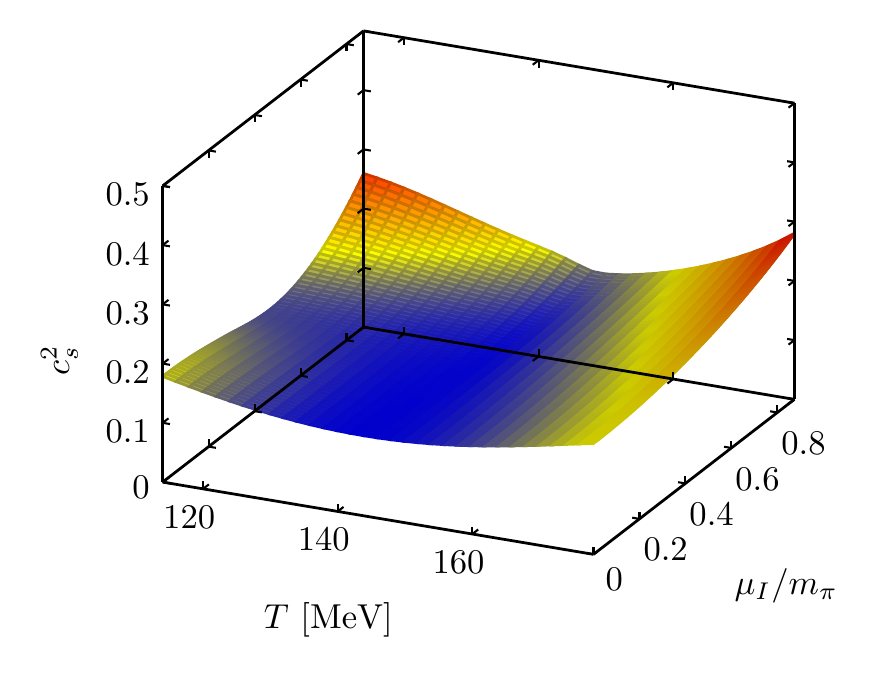}
 \caption{\label{fig:cs-Tneq0}
 Results for the isentropic speed of sound at $T\neq0$, obtained on the $N_t=8,\,10$ and 12
 lattices (from top left to bottom left) in the $T$-$\mu_I$ plane, excluding uncertainties.}
\end{figure}

\begin{figure}[t]
 \centering
\vspace*{-2mm}
\includegraphics[width=.48\textwidth]{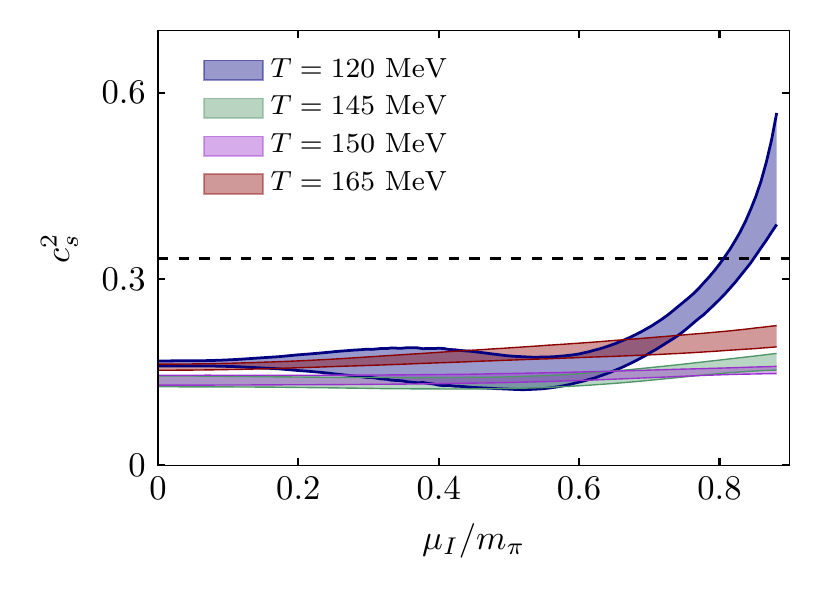}
\includegraphics[width=.48\textwidth]{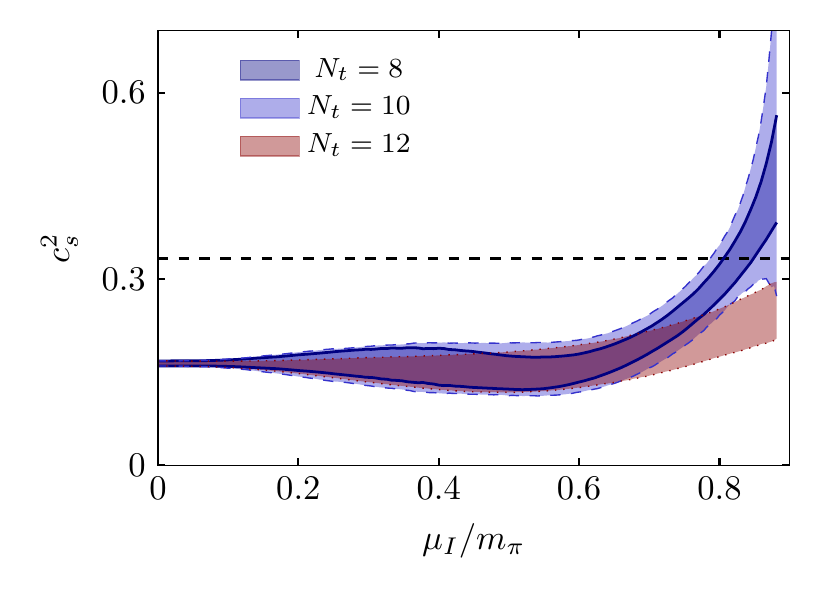}
 \caption{\label{fig:cs-Tneq0-slices}
 Results for the isentropic speed of sound at $T\neq0$ at different temperatures on the $8\times24^3$ lattices (left panel) and at a temperature of 120 MeV for $N_t=8$, 10 and 12 (right panel).}
\end{figure}

The results for the isentropic speed of sound at $T\neq0$ for the different lattices are shown in
Fig.~\ref{fig:cs-Tneq0}. For small temperatures, the speed of sound initially decreases slightly
in the vicinity of the BEC phase boundary, before it starts to rise within the BEC phase. For
$N_t=8$ and 10 it crosses the conformal bound around $\mu_I/m_\pi\approx0.75$ to 0.8, i.e., at
a somewhat larger $\mu_I$ value than at $T=0$. This is also visible in the
left panel
of Fig.~\ref{fig:cs-Tneq0-slices}, where we show the speed of sound obtained on the $N_t=8$ lattice
for different temperatures including the uncertainties. Contrary to the $T=0$ case, we do not observe
a maximum for $c_s^2$, likely due to the fact that it appears on the border or outside of our
$\mu_I$ range.

At this point we note that the speed of sound depends on the derivatives of $n_I$, which are
not well determined at the borders of our interpolation region at large $\mu_I$, as well as at
our largest and smallest temperatures (at $\mu_I=0$ it is determined by the parameterization
from Ref.~\cite{Borsanyi:2013bia}, see the lower left panel of Fig.~\ref{fig:eos-mui0}).
Consequently, the results for the speed of sound have to be taken with care beyond
$\mu_I/m_\pi\gtrsim0.85$. To allow the reader to scrutinize the uncertainties at low $T$ but
large $\mui$, we show in the right panel of Fig.~\ref{fig:cs-Tneq0-slices} $c_s^2$
including its uncertainties for the different $N_t$ at a temperature of $T=120$~MeV,
close to the lower border of the temperature range.
For $N_t=12$, the speed of sound does not reach the conformal bound,
but we can still observe an increase of $c_s$ towards larger values of $\mu_I$.
We interpret that as a shift of the maximum towards larger values of $\mu_I$ when we approach
the continuum. The question of the presence of the peak in the continuum limit can be answered
once the results on finer lattices and at larger $\mu_I$ become available.

\begin{figure}[t]
 \centering
\vspace*{-2mm}
\includegraphics[width=.48\textwidth]{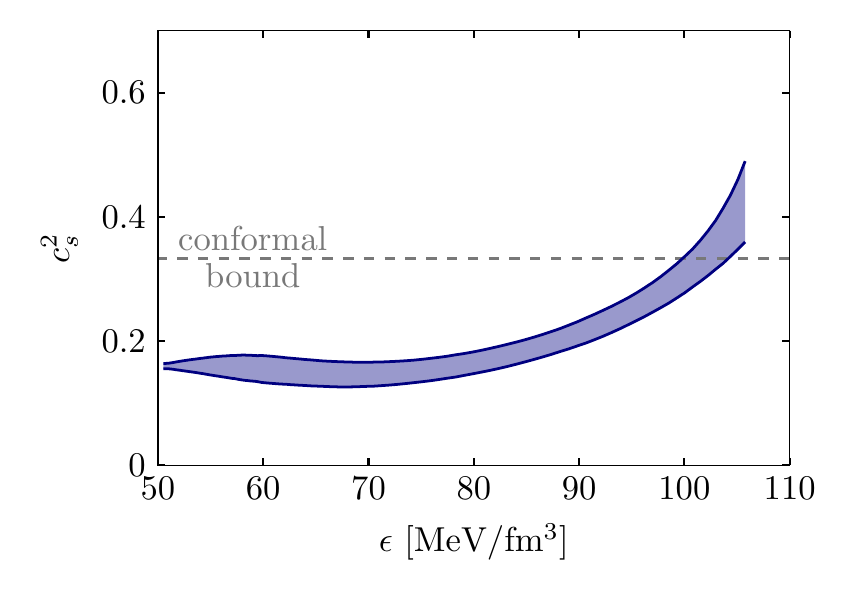}
\includegraphics[width=.48\textwidth]{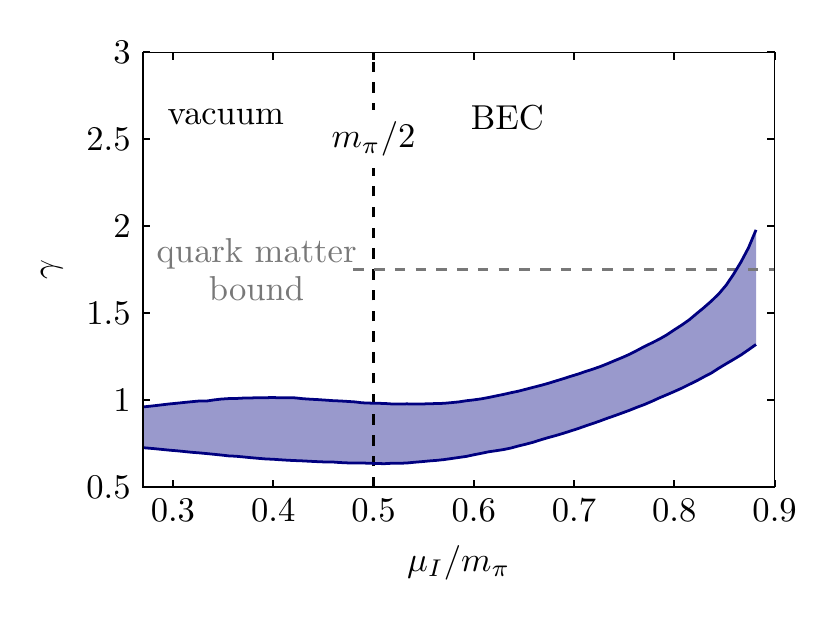}
 \caption{\label{fig:cs-Tneq0-gam}
 {\bf Left:} Results for the isentropic speed of sound versus the energy density at $T=122$~MeV
 obtained on the $8\times24^3$ lattices. Also shown is the conformal bound~\cite{Cherman:2009tw}
 (gray dashed line).
 {\bf Right:} Results for the polytropic index $\gamma$ at $T=122$~MeV. Also shown is the
 ``quark matter bound'' introduced in Ref.~\cite{Annala:2019puf} (gray dashed line).}
\end{figure}

Taking a closer look at the results from $N_t=8$, our most accurate results concerning the extraction
of $c_s$, albeit being furthest from the continuum, we show the squared speed of sound versus the
energy density and the polytropic index $\gamma$ versus $\mu_I$ for a small temperature of $T=122$~MeV
in Fig.~\ref{fig:cs-Tneq0-gam}.
Comparing the results for $c_s$ versus the energy density against those obtained at $T=0$,
shown in Fig.~\ref{fig:cs-T0-gam}, we observe that the speed of sound crosses the conformal
bound at about five times larger energy densities.
Contrary to what is seen at $T=0$, the polytropic index starts from a comparably small value
$\gamma=0.82(4)$, and does not increase directly
at the BEC phase boundary. The drastic increase towards larger values happens at
$\mu_I/m_\pi\approx0.7$ and it crosses the ``quark matter bound'' at around $\mu_I/m_\pi\approx0.85$.
Within our range of chemical potentials we see no sign of a maximum or the onset of a plateau.

\section{Discussion and conclusions}
\label{sec:conclusions}

In this paper we studied the QCD equation of state at nonzero temperatures and isospin chemical
potentials using first-principles lattice simulations at physical quark masses.
The results are shown in Figs.~\ref{fig:nt8-eos-dif-3d} and~\ref{fig:nt8-eos-3d}. We observe a
strong rise of the pressure within the phase of condensing charged pions (BEC phase)
at small temperature, which becomes milder when approaching the boundary of the BEC phase
with $T$. 
We mention that our results for the pressure might also be useful to constrain
the EoS at other chemical potentials. In our simulations the quark chemical potentials are set as $\mu_u=-\mu_d=\mu_I$ and $\mu_s=0$. Since flipping the sign of $\mu_d$ merely amounts to a phase change in the fermion determinant, it is simple to see that for the partition functions $\Z(\mu_u=-\mu_d)>\Z(\mu_u=\mu_d)$ holds. This implies that the pressure $\propto \log\Z$ at nonzero $\mu_I$ provides an upper bound for the pressure in a setup with equal light quark chemical potentials.

The interaction measure was found to initially rise
in the BEC phase, before it reaches a maximum and decreases. For small temperatures it
eventually becomes negative deep in the BEC phase, 
providing an explicit counter-example to general positivity arguments~\cite{Fujimoto:2022ohj}.
This effect diminishes as $T$ grows, and around the high-$T$ boundary of the BEC phase
the impact of $\mu_I$ is to shift the behavior of $I(T)$ to lower temperatures.
We also determine the QCD phase diagram in the
$T$-$\mu_I$ plane, which for $N_t=12$ is shown in Fig.~\ref{fig:nI-ph-diag}.

We have put a particular focus on the determination of the isentropic speed of sound $c_s$, for
which we show results both at zero and non-vanishing temperatures. At $T=0$, $c_s$ initially increases, crosses the conformal bound of $1/\sqrt{3}$
around $\mu_I/m_\pi\approx0.64$ and reaches a peak at $\mu_I/m_\pi\approx0.77$
with a maximum of $c_s^2\approx0.56$, see Eq.~\cref{eq:cs-peak} and Fig.~\ref{fig:cs-T0}.
It then decreases again and is expected to approach the conformal bound asymptotically from below.
To our knowledge, this is the first evidence for the explicit violation of this bound
in first principles QCD. 
We note  that similarly large $c_s$ values have also been observed at large
$\mu_B$ with the functional renormalization group approach in the point-like
approximation for four-quark interactions~\cite{Leonhardt:2019fua}.
A remnant of the peak in $c_s$ remains visible at low temperatures
(see Fig.~\ref{fig:cs-Tneq0}), where it is shifted towards larger values of $\mu_I$.
In Fig.~\ref{fig:phd} we also show a sketch of the region where
the speed of sound exceeds the conformal bound in the phase diagram.

Besides $c_s$,
we also computed the polytropic index $\gamma$, which has been
discussed as an indicator for the state of matter in the core of neutron
stars.
The results are shown in Figs.~\ref{fig:cs-T0-gam}
and~\ref{fig:cs-Tneq0-gam} (right panels). At $T=0$, the polytropic index starts from $\gamma=2$,
the value predicted by chiral perturbation theory~\cite{Son:2000xc}, and then drops below
$\gamma=1.75$, the ``quark matter bound'' introduced in~\cite{Annala:2019puf}, around
$\mu_I/m_\pi\approx 0.8$, eventually approaching the conformal value of $\gamma=1$
asymptotically. At $T\neq0$ the behavior is quite different. The polytropic index starts from
the $\mu_I=0$ value, around $\gamma=0.8$ to 1.0 -- see Fig.~\ref{fig:eos-mui0} bottom right --
from where it increases and approaches larger values of $\gamma\gtrsim1.75$ only around
$\mu_I/m_\pi\gtrsim 0.85$.

Our tabulated results for the EoS will be useful for comparison to low-energy models and effective theories of QCD. 
To this end, our data for the full EoS, including a code to compute the observables, are available with the published version of this paper.
Note that our calculations rely on three different lattice spacings, but owing to enhanced lattice artifacts at low temperatures for certain observables, we did not carry out a full
continuum extrapolation here.

We finally comment on the consequences of our finding of the excess of the speed
of sound over the conformal bound for the modeling of the EoS of neutron stars.
With increasing amount of data on the masses and radii of the observed neutron stars
in the Universe, several groups started to extract information on the QCD
EoS at nonzero baryon density using this experimental data. In these approaches,
the EoS is typically constrained at small and large densities from effective
hadronic models (e.g.~\cite{Gandolfi:2009fj,Tews:2012fj}) and perturbative
QCD (e.g.~\cite{Gorda:2018gpy}), respectively and then interpolated using a set of
basis functions
(e.g.~\cite{Hebeler:2013nza,Kurkela:2014vha,Annala:2017llu,Most:2018hfd} -- see also~\cite{Han:2021kjx,Han:2022rug} for non-parametric interpolations using neural networks).
Newer studies use a large set of different types of basic functions and millions
of different EoS
interpolations~\cite{Annala:2019puf,Somasundaram:2021clp,Annala:2021gom,Altiparmak:2022bke,Ecker:2022xxj,Marczenko:2022jhl}.
While most recent studies indicate that experimental constraints 
favour a stiff EoS with a speed of sound that exceeds the conformal
limit~\cite{Somasundaram:2021clp,Altiparmak:2022bke,Ecker:2022xxj,Marczenko:2022jhl,Han:2022rug},
it has often been considered as extreme for the EoS to develop large speeds of sound of
$c_s^2>0.5$ to 0.6 or even to have an EoS which exceeds the conformal bound. In our study
we provide direct evidence that an EoS with a speed of sound of this magnitude exists in QCD at small
temperatures. Thus, such conditions for the EoS of cold dense QCD matter are certainly not unrealistic.

\vspace*{2mm}\noindent
{\bf Acknowledgments:} \\
The authors are grateful to Szabolcs Bors\'anyi, Gergely Mark\'o, Guy Moore and Aleksi Vuorinen for
useful discussions and to K\'alm\'an Szab\'o for providing the parameterization for the
lattice QCD EoS at $\mui=0$. We also thank Prabal Adhikari, Jens Oluf Andersen and
Martin Mojahed for discussions and for providing the chiral perturbation theory data
from Ref.~\cite{Adhikari:2020kdn}.
The authors acknowledge support
by the Deutsche Forschungsgemeinschaft (DFG, German
Research Foundation) through the CRC-TR 211 “Strong-
interaction matter under extreme conditions” – project
number 315477589 – TRR 211. F.C. acknowledges the
support by the State of Hesse within the Research Cluster ELEMENTS
(Project ID 500/10.006). The authors
also gratefully acknowledge the Gauss Centre for Supercomputing e.V.
(\href{https://www.gauss-centre.eu}{\tt www.gauss-centre.eu})
for funding this project by providing computing time on the GCS Supercomputer
SuperMUC-NG at Leibniz Supercomputing Centre
(\href{https://www.lrz.de}{\tt www.lrz.de}).
Parts of the
computations in this work were performed on the GPU
cluster at Bielefeld University and at Goethe-HLR at
Goethe-University Frankfurt. We thank the computing staff
of both institutions for their support.

\appendix

\section{Zero density input to the EoS}
\label{app:mui0-eos}

\begin{table}[h!]
\begin{tabular}{cccccccc}
 $h_0$ & $h_1$ & $h_2$ & $f_0$ & $f_1$ & $f_2$ & $g_1$ & $g_2$ \\
 \hline
 0.1396(26) & -0.179(9) & 0.035(1) & 2.76(65) & 6.79(24) & -5.29(17) & -0.47(19) & 1.04(17)
\end{tabular}
\caption{\label{tab:eos-para-mui0}
Parameters for the parameterization of the $\mu_I=0$ EoS of Eq.~\eqref{eq:mi0-eos-def}, obtained from the reanalysis of the data of Ref.~\cite{Borsanyi:2013bia} as discussed in the text.}
\end{table}

\begin{figure}[t]
 \centering
 \includegraphics[width=7cm]{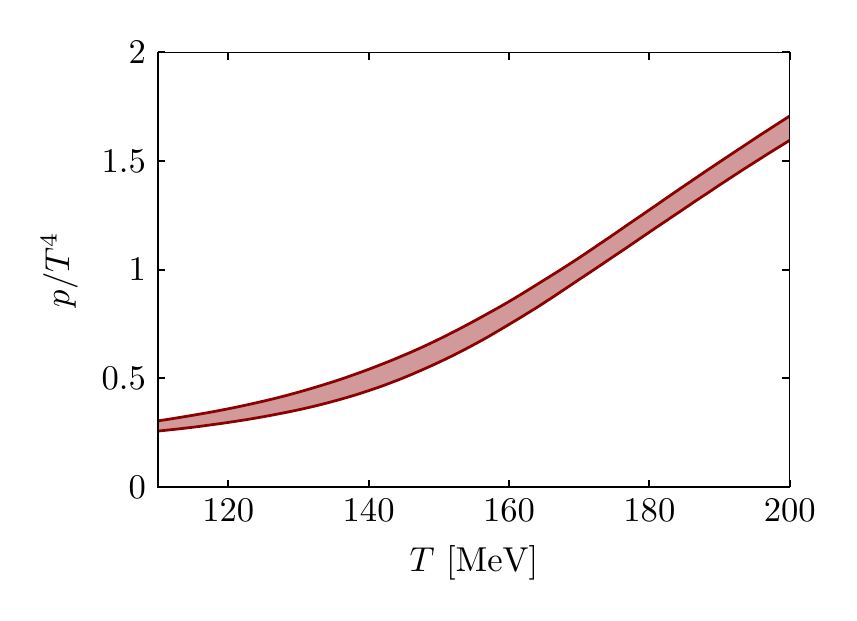}
 \includegraphics[width=7cm]{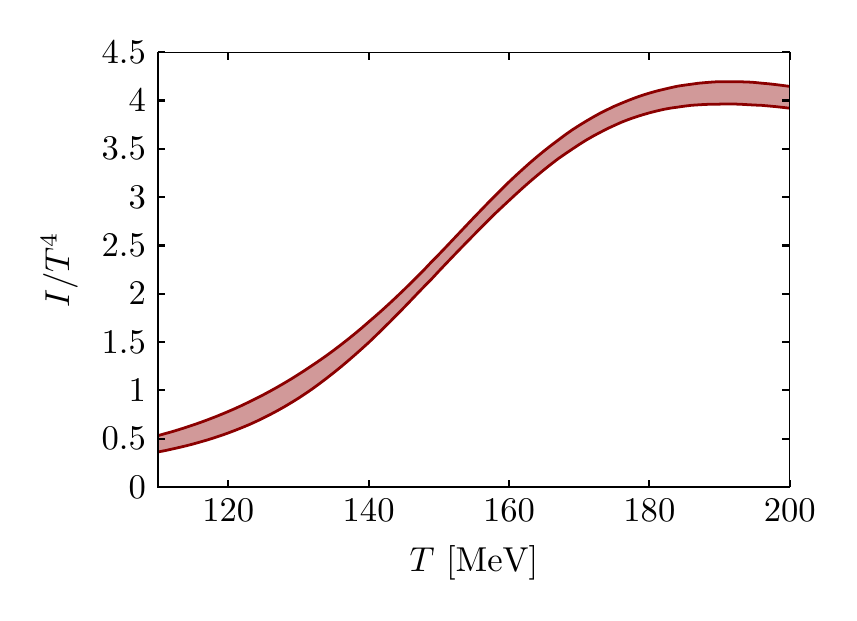} \\
 \includegraphics[width=7cm]{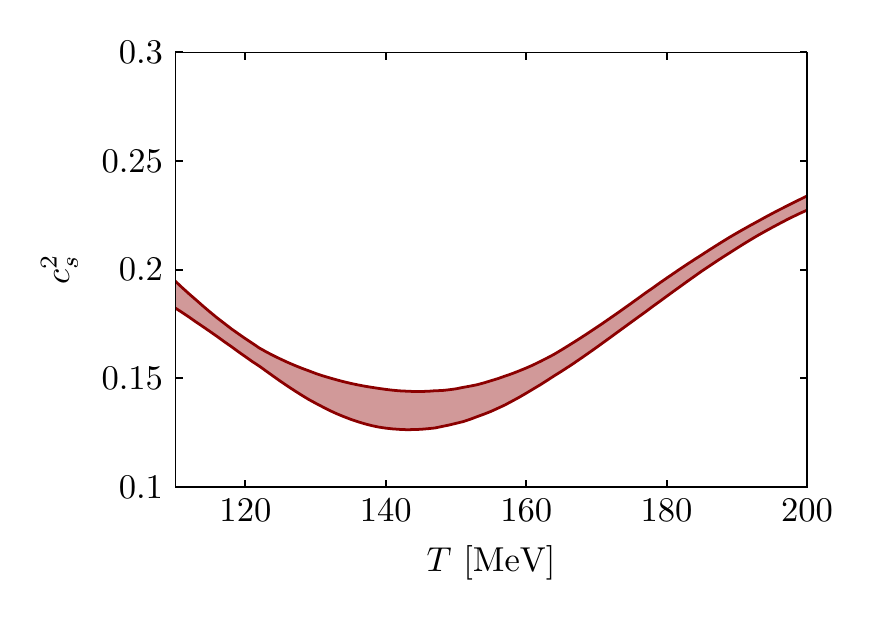}
 \includegraphics[width=7cm]{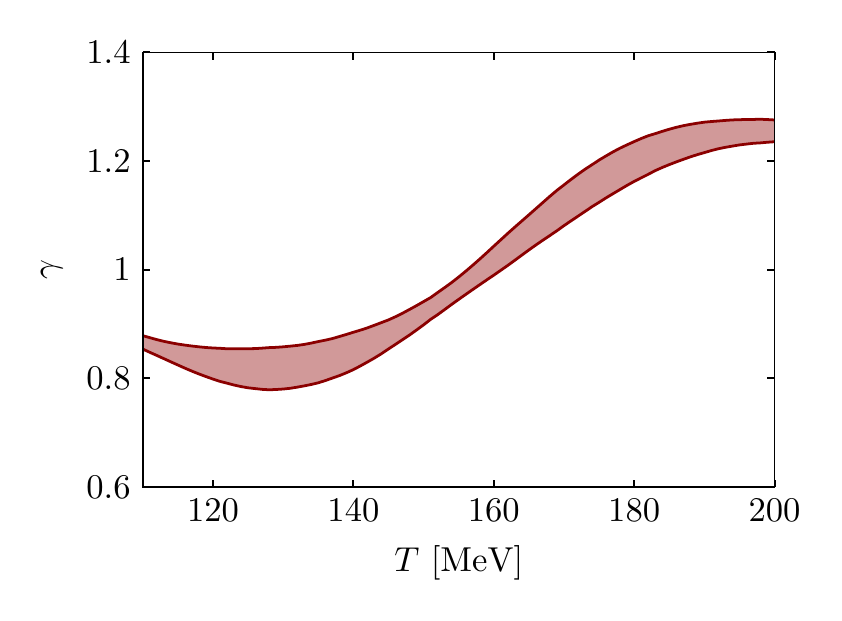}
 \caption{\label{fig:eos-mui0}
 Pressure, interaction measure, speed of sound squared and polytropic index
 in the temperature range
 relevant for this study, obtained from the parameterization at $\mu_I=0$
 of Ref.~\cite{Borsanyi:2013bia} discussed in the main text.}
\end{figure}

To compute the full EoS from the decompositions of Eq.~\eqref{eq:eos-decomp}, we need input
at $\mu_I=0$. Here we use the parameterization
\be
\label{eq:mi0-eos-def}
\frac{I(T,0)}{T^4} = e^{-h_1/t-h_2/t^2} \, \Big( h_0 +
\frac{f_0 \, \big[ \tanh(f_1\,t+f_2) + 1 \big] }{ 1 + g_1 t + g_2 t^2 } \Big) ,
\ee
with $t=T/(200\:\text{MeV})$, which has been employed in
Refs.~\cite{Borsanyi:2010cj,Borsanyi:2013bia}. Since we need to take the full correlations
between the parameters into account for the correct computation of the uncertainties of
derivatives and integrals, we use the parameters obtained from a
reanalysis
of the data from Ref.~\cite{Borsanyi:2013bia}. The resulting parameters are listed in
Tab.~\ref{tab:eos-para-mui0}. We note the slight differences in the parameters compared to
the ones obtained in Ref.~\cite{Borsanyi:2013bia}. These differences in the reanalysis
can be attributed to flat directions in parameter space, as already mentioned
in Ref.~\cite{Borsanyi:2013bia}, and do not lead to significant changes in the description
of the data and the curves for the thermodynamic observables.
For completeness, we show the results for the pressure,
the interaction measure, the squared speed of sound and the polytropic index
for this parameterization in the
parameter region relevant for this study in Fig.~\ref{fig:eos-mui0}.

\section{EoS from direct interaction measure calculations}
\label{app:eos-icomp}

An alternative to computing the EoS from the interpolation of the isospin density, is to
rewrite the interaction measure via the derivative of the
partition function with respect to the lattice scale, along similar lines as in Refs.~\cite{Allton:2003vx,Iida:2022hyy,Itou:2022ebw}. Using Eq.~\cref{eq:Ilat} and the
derivatives of the lattice parameters with respect to the lattice scale, one obtains
\be
 \label{eq:Ilat-lscale}
 I(T,\mu_I) = \frac{\partial \beta}{\partial \log a} \ev{S_g}_{T,\mu_I}
 - \sum_q \frac{\partial (am_q)}{a\,\partial \log a} \ev{\bar\psi\psi_q}_{T,\mu_I} \,,
\ee
where $\beta$ is the lattice coupling, $S_g$ the gauge action and $am_q$ the bare quark mass of flavour $q$ in lattice units. We note that all the quantities appearing in Eq.~\cref{eq:Ilat-lscale} need to
be renormalized properly, demanding, for instance, the knowledge of the quantities at $T=0$, but nonzero $\mu_I$. Another way to ensure a proper renormalization is to make use of the decomposition
of Eq.~\cref{eq:eos-decomp} and to compute only $\Delta I$ instead of $I$. Defining generically
$\Delta O = \ev{O}_{T,\mu_I} - \ev{O}_{T,0}$, we obtain
\be
 \label{eq:DIlat}
 \Delta I(T,\mu_I) = \frac{\partial \beta}{\partial \log a}
\Delta S_g - \sum_q \frac{\partial (am_q)}{a\,\partial \log a} 
\Delta \bar\psi\psi_q \,.
\ee
The remaining task is the computation of $\Delta S_g$ and $\Delta\bar\psi\psi_q$. 

\begin{figure}[ht]
 \centering
\vspace*{-2mm}
\includegraphics[width=.48\textwidth]{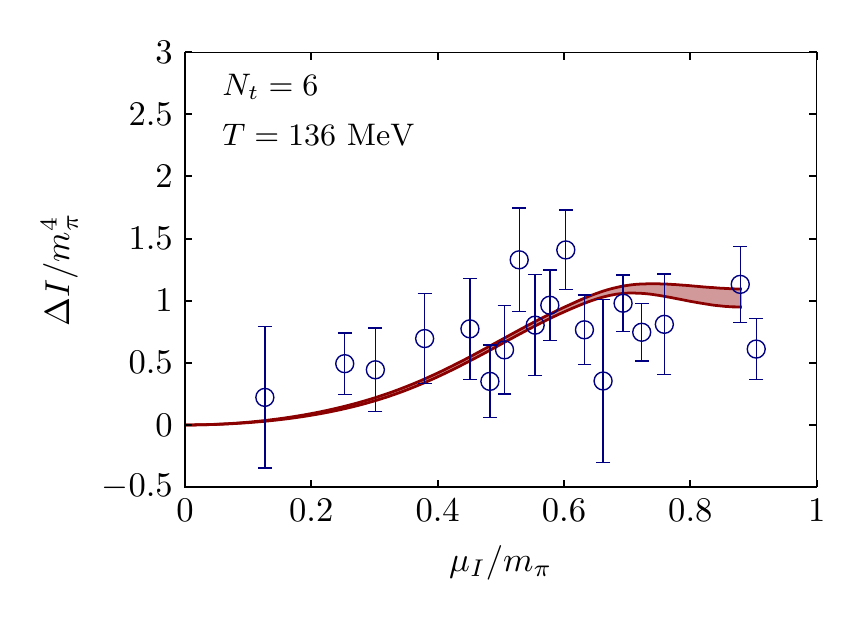}
\includegraphics[width=.48\textwidth]{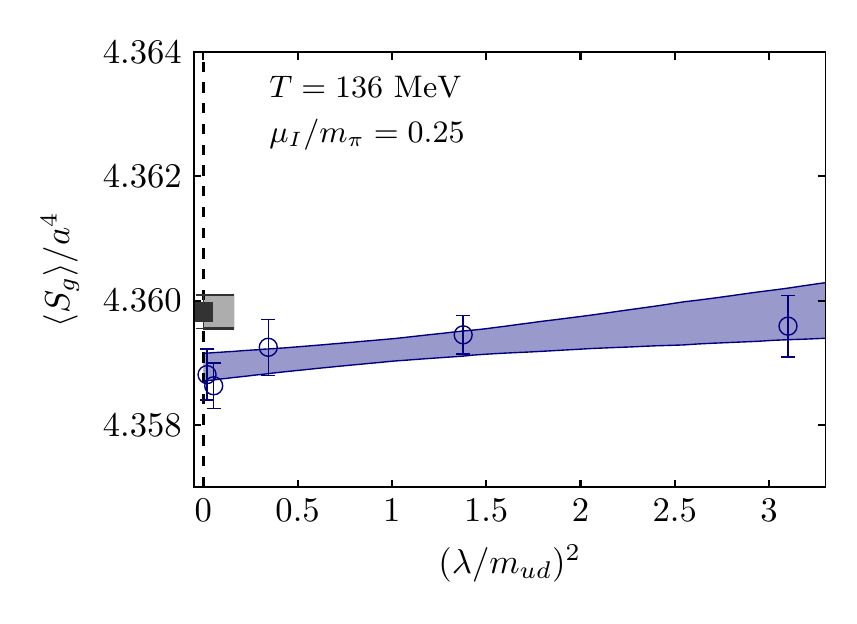}
 \caption{\label{fig:nt6-DI-icomp}
 {\bf Left:} Results for the interaction measure obtained from a direct computation
 explained in the text (blue points) and obtained from the two-dimensional interpolation
 of the isospin density (red band) on $24\times6$ lattices at a temperature of 136 MeV.
 {\bf Right:} $\lambda$-extrapolation (blue band) of the Symanzik improved gauge action
 (blue open circles) on a
 $24^3\times6$ lattice with $T=136$ MeV and $\mu_I/m_\pi=0.25$.
 We also show the result at $\mu_I=0$ for the same temperature
 (black filled box and black band) which needs to be subtracted from the $\lambda=0$
 result to obtain the contribution to $\Delta I$.}
\end{figure}

We show the results for $\Delta I(T,\mu_I)$ versus $\mu_I$ obtained for one particular
temperature on a set of $24^3\times6$ lattices in the left panel of
Fig.~\ref{fig:nt6-DI-icomp}. For comparison we also
show the result for $\Delta I$ obtained from the interpolation of the isospin density.
We note that the qualitative behavior of the two data sets is similar, at least
as far as comparison is possible due to the uncertainties for the direct computation,
which are at least an order of magnitude larger compared to the uncertainties of the results
obtained from the interpolation. The question is, where these large uncertainties orginate
from. The problem is the combination of the $\lambda$-extrapolations at $\mu_I\neq0$ in
combination with the subtraction of the $\mu_I=0$ value (despite the fact that such an extrapolation
is not necessary there). We show the $\lambda$-extrapolation of the Symanzik improved gauge action
in the right panel of Fig.~\ref{fig:nt6-DI-icomp} in comparison to the result at $\mu_I=0$ at
the same temperature, which when subtracted give $\Delta S_g$. As is evident from the plot,
the large uncertainties come from the subtraction of two quantities of similar magnitude,
so that relative uncertainties are enhanced by orders of magnitude compared to the uncertainties
of the individual quantities. This is particularly pronounced for the gauge action, but a similar
behavior is also seen for the quark condensates. We note, that such a subtraction is absent
for the isospin density, leading to way more accurate results for the quantity to interpolate
and, consequently, for the EoS.

\section[Isospin density at \texorpdfstring{$\lambda=0$}{l=0} and model-independent spline interpolations]{\boldmath Isospin density at \texorpdfstring{$\lambda=0$}{l=0} and model-independent spline interpolations}
\label{app:nI}

The basic data for the isospin density, Eq.~\cref{eq:nI-def}, is obtained at non-vanishing pion source, $\lambda\neq0$. For the extrapolations to $\lambda=0$ we use the improvement program from Refs.~\cite{Brandt:2017oyy,Brandt:2018bwq,Brandt:2018omg}. The resulting $\lambda$-extrapolations are basically flat and can be done using either a linear function in $\lambda^2$ or a constant. As an estimate for the systematic uncertainty associated with the extrapolation, we use the maximal deviation of the final result with either the extrapolation using the alternative (linear or constant), functional form or any of the two data-points at the lowest $\lambda$ values. Note, that this is particularly important for the extraction of the equation of state, since an underestimation of uncertainties might lead to unphysical fluctuations which significantly affect the interpolation using spline fits.

For the extraction of the equation of state we use an average over all possible cubic spline interpolations of the isospin density in the two-dimensional parameter space $(T,\,\mu_I)$ with less grid- than data points (i.e., spline fits), weighted with an estimator for the ``goodness'' of the spline fit. Note, that for our spline fits the positions of the spline nodepoints generically do not coincide with the positions of the data points. As already discussed in Ref.~\cite{Brandt:2016zdy}, this average for an observable $A$ (for instance the isospin density $n_I(T,\mu_I)$ for given values $T$ and $\mu_I$) can be written as
\be
\label{eq:spl-avg}
\big\llangle A \big\rrangle = \sum_{N_G} \int d^{N_x(N_G)}x
\,\, A(\vec{x}) \, \exp\big(-S_{\rm spl}(\vec{x},N_G)\big) \,.
\ee
Here $N_G$ is the total number of spline nodepoints and $N_x(N_G)$ the number of nodepoints which can be varied in the particular spline setup. Note that $N_x$ and $N_G$ do not need to be equivalent (but always $N_x\leq N_G$), since some of the nodepoint positions can be fixed. This is the case for the nodepoints at the lower end of the $n_I$-splines in the $\mu_I$ direction, for instance, which are kept at $\mu_I=0$. In Eq.~\eqref{eq:spl-avg}, $\vec{x}$ is the vector of (two-dimensional) nodepoint positions for the $N_x$ variable nodepoints. Note that typically the allowed range for the nodepoint values is restricted, as outlined below. The action $S_{\rm spl}(\vec{x},N_G)$ represents the estimate for the ``goodness'' of the spline fit. Possible choices have already been discussed in Ref.~\cite{Brandt:2017oyy}. As the basic action we use the Akaike information criterion~\cite{Akaike1973InformationTA} (see also Ref.~\cite{Akaike:1998zah}),
\be
\label{eq:Saic}
S_{\rm AIC} = 2 N_P + \chi^2 \,,
\ee
where $N_P$ is the number of parameters of the fit.

One of the major problems for any spline interpolation or spline fit is the possible appearance of oscillatory solutions, i.e., solutions with additional minima and maxima as the spline attempts to capture all of the datapoints. These solutions can in particular be triggered by statistical fluctuations of data points and are particularly problematic for the equation of state, since additional unphysical minima and maxima might have strong effects on quantities like the speed of sound. To suppress those solutions we include another term in the action, following the spirit of Ref.~\cite{Endrodi:2010ai}. The term signifies the stability of the spline solution under small variations of the nodepoints. The parameters of the spline $f_n$ (with $n=1,\ldots,N_P$) are given either by the value of the spline on one of the spline nodepoints or by the derivatives on the nodepoints of the spline boundaries, depending on the particular spline setup. If we vary the nodepoints slightly and have a stable, non-oscillatory spline solution, we expect those values to not change significantly. I.e., given a variation $\alpha$ of one of the nodepoints, here with nodepoint index $k$ and the variation in direction $i$,
\be
\label{eq:apl-vary}
(x^\alpha_k)_i = (x_k)_i + \epsilon \qquad \text{and} \qquad (x^\alpha_l)_j' = (x_l)_j \:\:\forall l\neq k,\,j\neq i ,
\ee
where $\epsilon$ is a small (not necessarily positive) number compared to the typical distance between two datapoints, we expect the spline parameters $f^\alpha_n$ to differ only slightly from the previous parameters $f_n$. The parameter variation with respect to the typical statistical uncertainties for the parameters can be estimated by
\be
\mathcal{D}^\alpha = \frac{1}{N_P}\sum_n \frac{| f^\alpha_n - f_n |}{\sigma(f_n)} \,.
\ee
Here $\sigma(f_n)$ is the statistical uncertainty of parameter $f_n$, obtained from applying the same spline fit to the individual bootstrap samples for the data points. For stable fits we expect $\mathcal{D}^\alpha$ to be a number not much larger than one, of course depending on the typical order of magnitude of the statistical uncertainties and the typical change of the value of two consecutive data points. To suppress unwanted oscillatory solutions in the sum of Eq.~\cref{eq:spl-avg}, we add to the action the average of $\mathcal{D}^\alpha$ over all possible spline variations $\alpha$,
\be
S_{\rm STAB} = \frac{1}{N_G} \sum_{\alpha} \mathcal{D}^\alpha \,,
\ee
so that the total action is given by
\be
S_{\rm spl} = S_{\rm AIC} + \gamma S_{\rm STAB}
\ee
with the tunable parameter $\gamma$.

The tunable parameters of the spline average outlined above concern the possible numbers of nodepoints in each direction, possible constraints on the nodepoint locations, the boundary conditions of the spline, i.e., for each spline boundary one derivative for a cubic spline, as used here, and the parameter $\gamma$ as well as the size of $\epsilon$. For the interpolation of the isospin density, we use three to five nodepoints in each direction and demand that always two data points reside between the outermost and the consecutive nodepoint on each border of the grid and at least one data point lies between two consecutive nodepoints in each direction. The application of these constraints to the spline grids is straightforward if the data points themselves form a rectangular grid. The outer nodepoints in $\mu_I$ direction are fixed at $\mu_I=0$ and on this whole boundary we impose $n_I=0$. To account for the isospin density being an uneven function in $\mu_I$, we also impose $\partial^2 n_I/(\partial \mu_I)^2|_{\mu_I=0}=0$. The second derivatives in all other directions have been kept as free parameters for the spline fit and the positions of the outer gridpoints are allowed to vary.

To efficiently perform the sum from Eq.~\cref{eq:spl-avg}, we use Monte-Carlo methods as proposed in Ref.~\cite{Brandt:2017oyy}. In particular, we employ a Metropolis algorithm with a symmetric proposal probability for changes in the spline nodepoints. For $\epsilon$ we have chosen a random number between a tenth of the distance between the two nearest data points in positive or negative direction in such a way, that the nodepoint remains between the two data points after variation. To tune the parameter $\gamma$, we have performed several runs starting from small values of $\gamma$ and monitored the resulting splines in the Markov chain. We stopped increasing $\gamma$ when we found that no significant oscillations leading to local minima/maxima structures between two nodepoints showed up in the final average over spline configurations. As the final value we choose $\gamma=10$. For our final results we have first tuned the number of nodepoints to the optimal value by performing 100 independent thermalisations allowing changes in the number of nodepoints and the nodepoint locations with respect to the data points. We then search for the spline with the lowest action and restrict ourselves to these number of nodepoints and the intervals in which the spline nodepoints reside with respect to the grid of data points. The final results are then obtained from 100 splines obtained by 20 independent runs where we vary the nodepoints in this constrained setup with 20000 thermalisation updates and with 5 spline configurations each, separated by 10000 spline updates.

\begin{table}
\centering
\begin{tabular}{c|cc|c|c}
 \hline
 $N_t$ & $N_G^{(T)}$ & $N_G^{(\mui)}$ & $T$-intervals [MeV] & $\mu_I$-intervals [MeV] \\
 \hline
 8 & 3 & 4 & $<114$, $[142,147]$, $>174$ & $[34,51]$, $[51,68]$, $>119$ \\
 \hline
 10 & 3 & 4 & $<114$, $[158,168]$, $>179$ & $[51,68]$, $[68,85]$, $>119$ \\
 \hline
 12 & 3 & 3 & $<114$, $[139,149]$, $>179$ & $[34,51]$, $>119$ \\
 \hline
\end{tabular}
\caption{Number of spline nodepoints in the $T$ and $\mui$ directions and the associated intervals in which the Monte-Carlo-generated nodepoints reside.}
\label{tab:spl-node}
\end{table}

\begin{figure}[ht]
 \centering
\vspace*{-2mm}
\includegraphics[width=.32\textwidth]{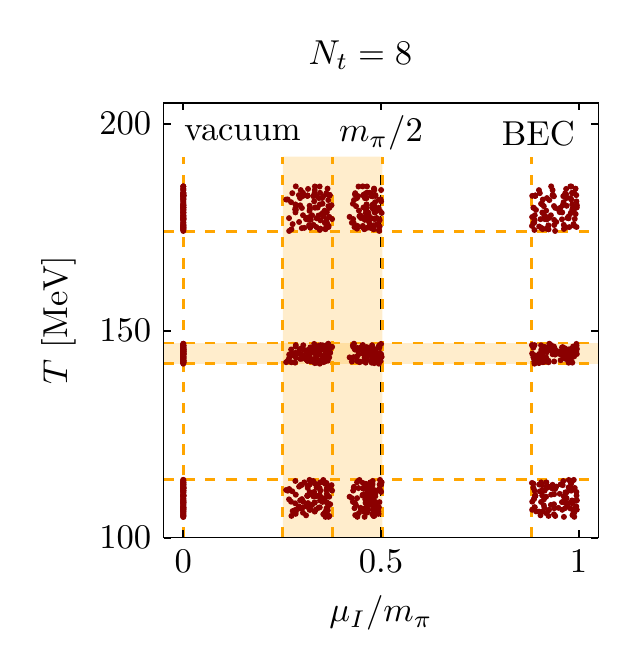}
\includegraphics[width=.32\textwidth]{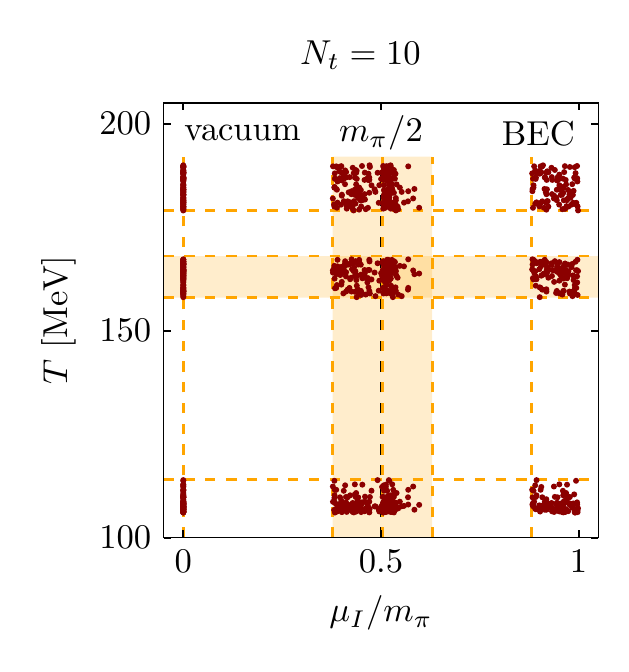}
\includegraphics[width=.32\textwidth]{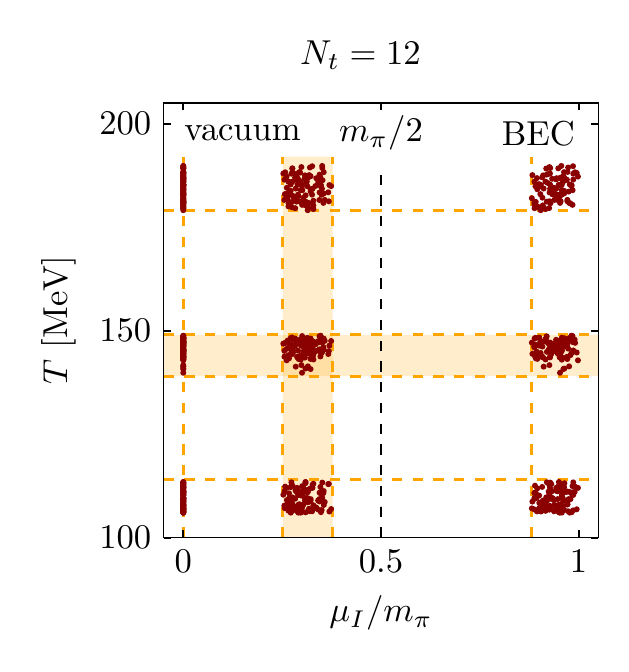}
 \caption{\label{fig:spl-nodes}
 Summary of the location of spline nodepoints in the final run. Note that the spline nodepoints can only be generated in certain regions of the parameter space, indicated by the accumulation of dots.}
\end{figure}

\begin{figure}[ht]
 \centering
\vspace*{-2mm}
\includegraphics[width=.48\textwidth]{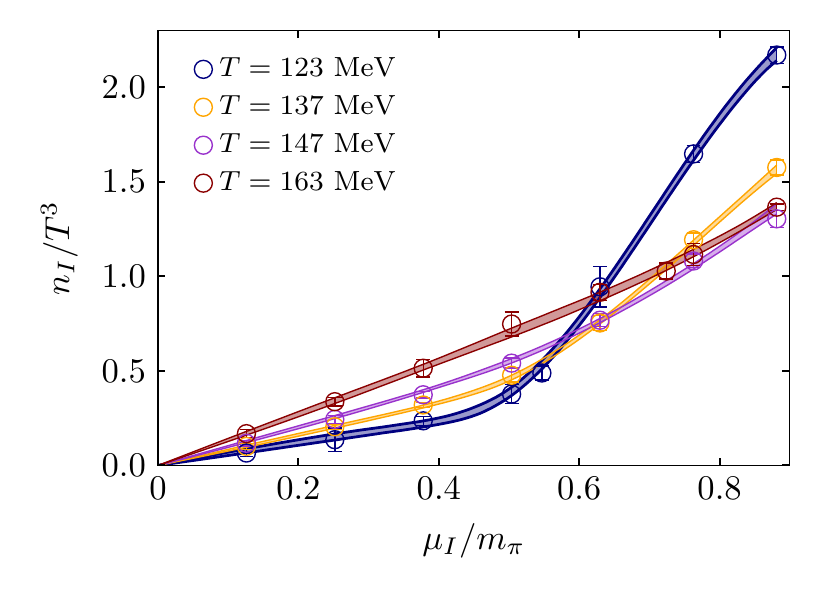}
\includegraphics[width=.48\textwidth]{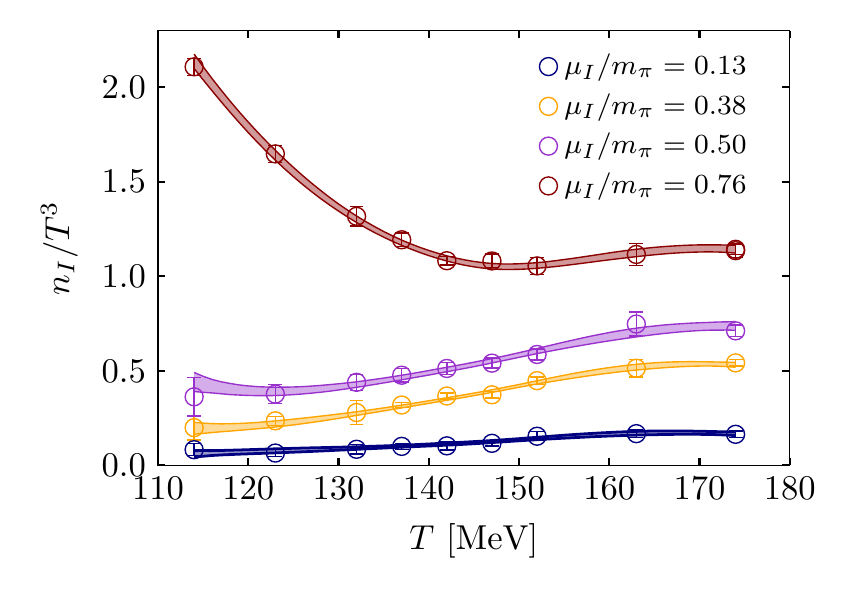}
 \caption{\label{fig:nt8-nI-spline}
 Comparison of the results for the isospin density $n_I$ versus $\mui$ for different
 temperatures (left) and $T$ for different $\mui$ (right) obtained from the simulations on
 $8\times 24^3$ lattices and the associated spline interpolation described in the text.}
\end{figure}

The final numbers of nodepoints in the different directions
$N_G=N_G^{(T)}\times N_G^{(\mui)}$, as well as the
intervals in which the nodepoint positions have been varied are given in Tab.~\ref{tab:spl-node}.
We show the location of the 100 final nodepoint sets used in the analysis in
Fig.~\ref{fig:spl-nodes}.
Note that the outer spline nodepoints can lie anywhere outside of the data point interval
and that the outer nodepoint of the lower boundary in $\mui$ direction has been held
fixed at $\mui=0$.
The resulting interpolation for $N_t=8$ lattices is shown for a set of temperatures in Fig.~\ref{fig:nt8-nI-spline}. The uncertainties include the uncertainty due to the individual data points (computed using the bootstrap procedure with 1000 samples) and from the Monte-Carlo over spline interpolations.
As a comment concerning the $T=0$ splines, in this case we are speaking about a one-dimensional spline interpolation, which is  rather well behaved and uncritical concerning the spline Monte-Carlo. In this case, the number of nodepoints and their location has been allowed to vary freely up to the maximal possible number of nodepoints given the number of available data points with the matching to chiral perturbation theory as discussed in section~\ref{sec:cs-t0dis}.

\section{Simulation Details}
\label{app:sim-detail}

We provide the run parameters for the simulations at $T\neq0$ and $T=0$ in
Tab.~\ref{tab:sim-paras}. For each of the mentioned parameter
values we have simulated at up to five different values of the pion
source parameter $\lambda$ with $1.4>\lambda/m_{ud}\gtrsim0.05$. At $T=0$ we have used
a fixed number of three different $\lambda$ values. For each of these
parameter sets we generated between 500 and 2000 trajectories, measuring observables
(here the isospin density $n_I$) on every fifth configuration. Typically we use two to three
independent chains to acquire the full statistics. We estimate autocorrelations using
the integrated autocorrelation time of the plaquette expectation value. The values we obtain
on the different ensembles are mostly between 5 and 10 in molecular dynamics units,
with the tendency to large values for smaller temperatures and $\lambda$-values, as well
as for larger values of $\mu_I$ and $N_t$. While this indicates that typically two consecutive
configurations are correlated, the autocorrelation times obtained for $n_I$, the main observable of our study, are much smaller. In rare cases we observed autocorrelation times of the order of 20 in molecular dynamics units, so that 4 consecutive measurements are correlated. Generically, we have checked the error analysis by using binning prior to the bootstrap procedure, but did not observe a significant dependence of the uncertainties on the binsize. Furthermore, in the improved $\lambda$-extrapolation, values at different $\lambda$ are combined using a linear function to extract
the $\lambda=0$ result, averaging out the fluctuations of the individual simulation points.

\begin{table}
\centering
\begin{tabular}{cc|cc|c}
 \hline
 $N_t$ & $N_s$ & $T [\text{MeV}]$ & $\beta$ & $a \mui$ \\
 \hline
 8 & 24 & 114 & 3.5500 & 0.019 0.037 0.056 0.075 0.094 0.113 0.131 \\
8 & 24 & 123 & 3.5750 & 0.018 0.034 0.052 0.070 0.075 0.087 0.105 0.122 \\
8 & 24 & 132 & 3.6000 & 0.016 0.032 0.048 0.065 0.075 0.081 0.098 0.113 \\
8 & 24 & 137 & 3.6120 & 0.016 0.031 0.047 0.063 0.079 0.094 0.109 \\
8 & 24 & 142 & 3.6250 & 0.015 0.030 0.045 0.060 0.075 0.076 0.091 0.106 \\
8 & 24 & 147 & 3.6370 & 0.015 0.029 0.044 0.058 0.073 0.088 0.102 \\
8 & 24 & 152 & 3.6500 & 0.014 0.028 0.042 0.056 0.071 0.075 0.085 0.098 \\
8 & 24 & 163 & 3.6750 & 0.013 0.026 0.039 0.053 0.066 0.075 0.079 0.092 \\
8 & 24 & 174 & 3.7000 & 0.012 0.024 0.037 0.049 0.062 0.074 0.075 0.086 \\
 \hline
 10 & 28 & 114 & 3.6250 & 0.015 0.030 0.045 0.060 0.075 0.090 0.105 \\
10 & 28 & 122 & 3.6500 & 0.014 0.028 0.042 0.056 0.070 0.084 0.098 \\
10 & 28 & 130 & 3.6750 & 0.013 0.026 0.039 0.052 0.065 0.078 0.091 \\
10 & 28 & 139 & 3.7000 & 0.012 0.024 0.037 0.049 0.061 0.073 0.086 \\
10 & 28 & 149 & 3.7250 & 0.011 0.023 0.034 0.046 0.057 0.069 0.080 \\
10 & 28 & 158 & 3.7500 & 0.011 0.022 0.032 0.043 0.054 0.065 0.075 \\
10 & 28 & 168 & 3.7750 & 0.010 0.020 0.030 0.040 0.051 0.061 0.071 \\
10 & 28 & 179 & 3.8000 & 0.010 0.019 0.029 0.038 0.048 0.057 0.067 \\
 \hline
 12 & 36 & 114 & 3.6900 & 0.012 0.025 0.037 0.050 0.063 0.075 0.088 \\
12 & 36 & 122 & 3.7200 & 0.011 0.023 0.034 0.046 0.058 0.069 0.081 \\
12 & 36 & 132 & 3.7500 & 0.010 0.021 0.032 0.043 0.054 0.064 0.075 \\
12 & 36 & 140 & 3.7750 & 0.010 0.020 0.030 0.040 0.051 0.060 0.071 \\
12 & 36 & 149 & 3.8000 & 0.009 0.019 0.028 0.038 0.048 0.057 0.067 \\
12 & 36 & 158 & 3.8250 & 0.009 0.018 0.027 0.036 0.045 0.054 0.063 \\
12 & 36 & 167 & 3.8500 & 0.008 0.017 0.025 0.034 0.043 0.051 0.060 \\
12 & 36 & 179 & 3.8800 & 0.008 0.016 0.023 0.032 0.040 0.048 0.056 \\
 \hline
32 & 24 & 29 & 3.5500 & 0.067 0.074 0.082 0.093 0.112 0.130 \\
 \hline
48 & 32 & 27 & 3.6700 & 0.053 0.059 0.066 0.080 0.092 0.106 \\
 \hline
\end{tabular}
\caption{Summary of simulation parameters used for the determination of the EoS.}
\label{tab:sim-paras}
\end{table}

%\bibliographystyle{JHEP}
%\bibliography{eosb}

\providecommand{\href}[2]{#2}\begingroup\raggedright\endgroup

\end{document}